\documentclass[aps,pra,twocolumn,showpacs,floatfix,superscriptaddress]{revtex4-1}
\usepackage{amsmath,amssymb,amsfonts,bbm,graphicx,hyperref,times}
\usepackage{amsthm}
\usepackage{amsfonts}
\usepackage{microtype}
\usepackage{mathrsfs}
\usepackage{placeins}
\usepackage{bbm}
\usepackage{booktabs}
\usepackage{color}

\newcommand{\blambda}{{\boldsymbol \lambda}}

\newcommand{\E}{\mathcal{E}_\epsilon}
\newcommand{\emaxU}{\epsilon_{\sf max}^{(1)}}
\newcommand{\emaxD}{\epsilon_{\sf max}^{(2)}}
\newcommand{\Tr}{{\rm Tr}}
\newcommand{\bra}[1]{\left\langle #1 \right\vert}
\newcommand{\ket}[1]{\left\vert #1 \right\rangle}

\newtheorem{lemma}{Lemma}
\newtheorem{proposition}{Proposition}
\newtheorem{corollary}{Corollary}
\newtheorem{criterion}{Criterion}
\newcommand{\proofend}{\hfill\blacksquare\\\medskip }
\renewcommand{\proof}[1]{{\noindent\bf Proof } #1 $\proofend$}
\begin{document}
\title{Detecting quantum non-Gaussianity via the Wigner function}
\author{Marco G. Genoni}
\affiliation{QOLS, Blackett Laboratory, Imperial College London, London SW7 2BW, UK}
\author{Mattia L. Palma}
\affiliation{Dipartimento di Fisica, Universit\`a degli Studi di Milano, I-20133 Milano, Italy}
\affiliation{QOLS, Blackett Laboratory, Imperial College London, London SW7 2BW, UK}
\author{Tommaso Tufarelli}
\affiliation{QOLS, Blackett Laboratory, Imperial College London, London SW7 2BW, UK}
\author{Stefano Olivares}
\affiliation{Dipartimento di Fisica, Universit\`a degli Studi di Milano, I-20133 Milano, Italy}
\author{M. S. Kim}
\affiliation{QOLS, Blackett Laboratory, Imperial College London, London SW7 2BW, UK}
\author{Matteo G. A. Paris}
\affiliation{Dipartimento di Fisica, Universit\`a degli Studi di Milano, I-20133 Milano, Italy}
\begin{abstract}
We introduce a family of criteria to detect quantum non-Gaussian states of a
harmonic oscillator, that is, quantum states that can not be expressed as
a convex mixture of Gaussian states. In particular we prove that, 
for convex mixtures of Gaussian states, the value of the Wigner
function at the origin of phase space is bounded from below by a non-zero positive
quantity, which is a function only of the average number of excitations
(photons) of the state. As a consequence, if this bound is violated then
the quantum state must be quantum non-Gaussian. We show that this criterion
can be further generalized by considering additional Gaussian operations
on the state under examination. We then apply these criteria to various
non-Gaussian states evolving in a noisy Gaussian channel, proving
that the bounds are violated for high values of losses, and thus also
for states characterized by a positive Wigner function.  \end{abstract}
\pacs{03.65.Ta, 42.50.Dv}
\maketitle
\section{Introduction}
Several criteria to detect non-classicality of quantum states of a 
harmonic oscillator have been introduced, mostly 
based on phase-space distributions
\cite{Wig32,Gla63,Sud63,CTL91,CTL92,Arv97,Arv98,Mar01,Par01,Dod02,Ken04}, 
ordered moments \cite{Shc05,Kie08,Vog08}, or on information-theoretic arguments
\cite{Sim00,Dua00,Mar02,Gio10,Fer12,Geh12,Por12}.  At the same time, an
ongoing research line addresses the characterization of quantum states
according to their Gaussian or non-Gaussian character
\cite{nonGHS,nonGRE,nonGL,barbieri10,Rad11,Jez11,Jez12,predojevic12,Rad13,Dau10}, 
and a question arises on whether those two
different hierarchies are somehow linked each other. \par
\begin{figure}[t!]
\includegraphics[width=0.95\columnwidth]{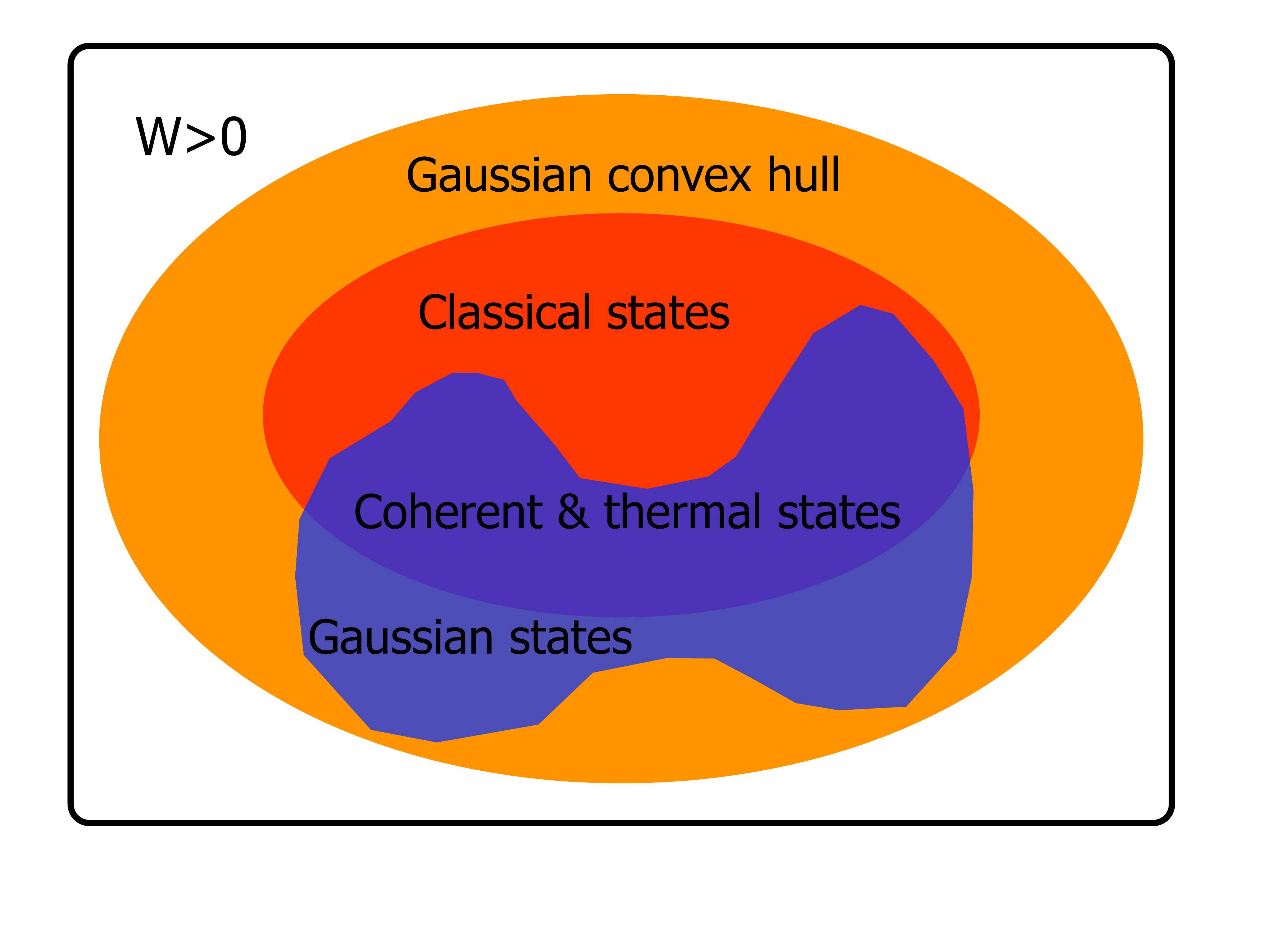}
\caption{Venn diagram description for continuous-variable quantum states
with positive Wigner function. The quantum states can be divided in two
sets: {\em quantum non-Gaussian} states and states belonging to the
Gaussian convex hull. The latter trivally includes (Glauber) classical
states and Gaussian states.  \label{f:PosWigner}} \end{figure}

 
As a matter of
fact, if we restrict our attention to pure states, Hudson's theorem
\cite{hudson74,soto83} establishes that the border between Gaussian and
non-Gaussian states coincides exactly with the one between states with
positive and negative Wigner functions. However, if we move to mixed
states, the situation gets more involved. Attempts to extend 
Hudson's theorem have been made, by looking at upper bounds on
non-Gaussianity measures for mixed states having positive Wigner
function \cite{mandilara09}. In this framework, by focusing on states
with positive Wigner function, one can define an additional border
between states in the {\em Gaussian convex hull} and those in the
complementary set of {\em quantum non-Gaussian states}, that is, states
that can not be expressed as mixtures of Gaussian states. The situation
is summarized in Fig. \ref{f:PosWigner}:  the definition of the
Gaussian convex hull generalizes the notion of Glauber's
non-classicality \cite{titulaer65}, with coherent states replaced by
generic pure Gaussian states, i.e. squeezed coherent states.
\par
Quantum non-Gaussian states with positive Wigner function are not useful 
for quantum computation \cite{mari12,veitch13}, and are not necessary for 
entanglement distillation, {\em e.g.} the non-Gaussian entangled resources
used in \cite{heersink06} are mixtures of Gaussian states. On the other
hand, they are of fundamental interest for quantum information 
and quantum optics. In particular, since no negativity of the
Wigner function can be detected for optical losses higher than $50\%$ 
\cite{cahill69} (or equivalently, for detector efficiencies  below $50\%$) 
criteria able to detect quantum non-Gaussianity are needed in order to 
certify that a {\em highly non linear} process (such as Fock state generation, 
Kerr interaction, photon addition/subtraction operations or conditional 
photon number detections) has been implemented in a noisy environment, even 
if no negativity can be observed in the Wigner function.
\par
Different measures of non-Gaussianity for quantum states have been
proposed \cite{nonGHS,nonGRE,nonGL}, but these cannot
discriminate between quantum non-Gaussian states and mixtures of
Gaussian states. An experimentally friendly criterion for quantum
non-Gaussianity, based on photon number probabilities, has been 
introduced \cite{Rad11}, and then employed
in different experimental settings to prove the generation of quantum
non-Gaussian states, such as heralded single-photon states
\cite{Jez11}, squeezed single-photon states \cite{Jez12} and Fock
states from a semiconductor quantum dot \cite{predojevic12}. 
\par
In this paper we introduce a family of criteria which are able to detect
quantum non-Gaussianity for single-mode quantum states of a harmonic
oscillator based on the Wigner function. 
As we already pointed out, according to Hudson's theorem, the only pure states having a
positive Wigner function are Gaussian states. One can then wonder if any
bound exists on the values that the Wigner function of convex mixtures
of Gaussian states can take. By following this intuition we 
present several bounds on the values of the Wigner function for
convex mixtures of Gaussian states, consequently defining a class of sufficient
criteria for quantum non-Gaussianity. 
\par
In the next section we will introduce some notation and the preliminary
notions needed for the rest of the paper. In Sec. \ref{s:criteria} we
will prove and discuss our Wigner function based criteria for quantum
non-Gaussianity and in Sec.  \ref{s:examples} we will prove their
effectiveness by considering different families of non-Gaussian states
evolving in a lossy (Gaussian) channel.  We will conclude the paper in
Sec. \ref{s:conclusions} with some remarks.
\section{Preliminary notions} \label{s:preliminary}
Throughout the paper we
will use the quantum optical terminology, where excitations of a quantum
harmonic oscillator are called photons. All the results can be naturally
applied to any bosonic continuous-variable (CV) system. We will
consider a single mode described by a mode operator $a$, satisfying the
commutation relation $[a,a^\dag] = \mathbbm{1}$. A quantum state
$\varrho$ is fully described by its
characteristic function \cite{cahill69}
\begin{align}
\chi[\varrho](\gamma) = \Tr[\varrho D(\gamma)]\:,
\end{align}
where $D(\gamma) = \exp\{\gamma a^\dag - \gamma^* a \}$ represents the displacement
operator. In addition, the quantum state $\varrho$ can be fully described by the 
Fourier transform of the characteristic function, {\em i. e.} the Wigner 
function \cite{cahill69} 
\begin{align}
W[\varrho](\alpha) = \int \frac{d^2 \gamma}{\pi^2} e^{\gamma^* \alpha - \gamma \alpha^*} 
\chi[\varrho](\gamma)\:.
\end{align}
A state is defined to be {\em Gaussian} if and only if its Wigner function 
(or equivalently its characteristic function) is Gaussian. All
single-mode Gaussian states can be expressed as $$\varrho= D(\alpha)
S(\xi) \nu_\beta S^\dag(\xi) D^\dag(\alpha)\,,$$ where 
$
S(\xi) = \exp\left\{ \frac12 \xi (a^\dag)^2 - \frac12 \xi^* a^2 \right\}
$,
is the squeezing operator, and 
$\nu_\beta = e^{-\beta a^\dag a}/\Tr[ e^{-\beta a^\dag a}]$ is a 
thermal state ($\alpha, \xi \in \mathbbm{C}$ and $\beta>0$). 
Pure Gaussian states can be written as $|\psi_{\sf G}\rangle =
D(\alpha) S(\xi)|0\rangle$, and, according to Hudson's theorem \cite{hudson74,soto83},
they are the only pure states having a positive Wigner function. 
Together with that of a Gaussian state, one can define the concept of {\em Gaussian
map}: a quantum (completely positive) 
map is defined Gaussian iff it transforms Gaussian states
into Gaussian states. All unitary Gaussian maps can be expressed as
$U_{\sf G} = \exp\{- i H_{\sf bil} t\}$, and they correspond to
Hamiltonian operators $H_{\sf bil}$ at most bilinear in the mode
operators. Similarly, a generic Gaussian map can be decomposed as a
Gaussian unitary acting on the system plus an ancilla (the latter
prepared in a Gaussian state), followed by partial tracing over the
ancillary mode \cite{eisert_wolf}.  
\par
Another complete description of a CV quantum state $\varrho$ may be
given in terms of the so-called
$P$-function $P[\varrho](\alpha)$ \cite{cahill69}, defined 
implicitly via the formula
\begin{align}
\varrho = \int d^2\alpha\: P[\varrho](\alpha) 
|\alpha\rangle\langle \alpha|\:, \label{eq:pfunction}
\end{align}
where $|\alpha\rangle = D(\alpha) |0\rangle$ represents a 
coherent state. According to Glauber a state 
is {\em non-classical} iff its $P$-function is not a proper probability 
distribution, {\em e.g.} the $P$-function is more singular than a 
Dirac-delta function. Note that the negativity of the Wigner function 
is a more restrictive definition of non-classicality: there exists 
non-classical states having a positive Wigner function ({\em e.g.} squeezed states),
while all the states having a non-positive Wigner function are
non-classical according to Glauber.
\par
In a similar spirit as in Glauber's approach to 
non-classicality, in this paper we study the concept of 
{\em quantum non-Gaussian} states. These are defined as 
follows. The Gaussian convex hull is the set of states
\begin{align}\label{eq:Ghull}
\mathcal{G} = \left\{ \varrho \in \mathcal{H} \: \lvert \: \varrho = \int d\blambda \: p(\blambda) \: |\psi_{\sf G}(\blambda)\rangle\langle\psi_{\sf G}(\blambda)|  \right\} \:,
\end{align}
where $\mathcal{H}$ denotes the Hilbert space of continuous-variable quantum states,
$p(\blambda)$ is a proper probability 
distribution and $|\psi_{\sf G}(\blambda)\rangle$ are pure 
Gaussian states, i.e. , in the single mode case, 
squeezed coherent states identified by the set of 
parameters $\blambda\equiv \{\alpha,\xi\}$. 
Since Gaussian states do not form a convex set, the set in Eq.~\eqref{eq:Ghull} 
includes states which are not Gaussian. Moreover any mixed Gaussian
state can be written as a weighted sum of pure Gaussian states, and hence 
the set above also includes convex mixtures of mixed Gaussian states. 
\par
The definition of {\em quantum non-Gaussianity} naturally follows:
\par\noindent
{\bf Definition}. {\em A quantum state $\varrho$ is quantum non-Gaussian 
iff it is not possible to express it as a convex mixture of Gaussian
states, that is iff $\varrho \notin \mathcal{G}$}.
\par
As illustrated in Fig. \ref{f:PosWigner}, the border here defined dividing
quantum non-Gaussian states and mixtures of Gaussian states falls in
between the border dividing classical and non-classical states, and the
one which divides states with positive and non-positive Wigner functions. The
importance of such a further distinction is evident if we note that all
states in ${\cal G}$ can be prepared through a combination of Gaussian
operations and classical randomization. 
On the contrary, if $\varrho\notin{\cal G}$ then a highly {\em non-linear} process
(due to a non-Gaussian operation or measurement) had necessarily taken place
in the generation the quantum state $\varrho$. While the negativity of
the Wigner function is always sufficient to certify it, more elaborated
criteria, as those elaborated in this paper, are needed in order to 
detect such a characteristic when
quantum states exhibit a positive Wigner function.
\section{Criteria to detect quantum non-Gaussianity}\label{s:criteria}
In order to find criteria for the detection of
quantum non-Gaussian states, we follow the
intuition given by Hudson's theorem for pure Gaussian states. We will
focus on lower bounds on the values taken by the Wigner function of
states which belong to the Gaussian convex hull $\mathcal{G}$.  In this
section we present our main findings as one lemma leading to two final
propositions and two additional corollaries. The `quantum
non-Gaussianity criteria', derived directly from these results, are
presented at the end of the section.
\begin{lemma}[Lower bound on the Wigner 
function at the origin of phase space for a pure Gaussian state]\label{l:boundpure}
For any given pure single-mode Gaussian state $|\psi_{\sf G}\rangle$, 
the value of the Wigner function at the origin of the phase space is bounded from below as 
\begin{align}
W[|\psi_{\sf G}\rangle\langle \psi_{\sf G}|](0) \geq \frac2\pi \exp\{ -2 n (1+n)\} \:, \label{eq:boundpure}
\end{align}
where $n=\langle \psi_{\sf G}| a^\dag a |\psi_{\sf G}\rangle$.
\end{lemma}
\proof{A generic pure single-mode Gaussian state can be always written as $|\psi_{\sf G}\rangle =
D(\alpha) S(\xi) |0\rangle$, where $\alpha=|\alpha|e^{i\theta}$ and $\xi=r e^{i\phi}$ ($r>0$) 
are two complex numbers. We can thus write the Wigner function evaluated in zero as 
\begin{align}
W\left[\ket{\psi_g}\bra{\psi_g}\right]&(0)=\frac{2}{\pi}\exp\left\{-2|\alpha|^2 \left[\cosh{2r} \right.
\right.\nonumber \\
&\left.\left. -\cos{(2\theta+\phi)}\sinh{2r} \right]  \right\} . 
\label{eq:wigGausszero}
\end{align}
Our goal is to minimize the value of the Wigner function or equivalently to maximize
the function
\begin{align}
f(\alpha,\xi)&=2|\alpha|^2 \left(\cosh{2r}-\cos{(2\theta+\phi)}\sinh{2r} \right).
\end{align}
A first maximization is obtained by considering
\begin{align}
2\theta+\phi=\pi + 2k\pi\:\:\:{\rm with}\:\: k\in \mathbbm{N}\:, \label{eq:phases}
\end{align}
which yields
\begin{align}
f(\alpha,\xi) \leq 2|\alpha|^2 e^{2 r} = 2n_d\left( 2n_s + 1 +2\sqrt{n_s(1+n_s)}\right).
\label{eq:max1}
\end{align}
In the last equation we introduced the displacement and squeezing photon numbers, $n_d = |\alpha|^2$ and $n_s =\sinh^2 r$, and we used the formula ${\rm arcsinh}(x)=\log(x+\sqrt{1+x^2})$. Note that these two parameters obey
$$
n=\langle \psi_{\sf G} | a^\dag a | \psi_{\sf G}\rangle =n_d + n_s,
$$ 
where $n$ is the average photon number of the state $| \psi_{\sf G}\rangle$. We can thus express the right hand side (rhs) of Eq. (\ref{eq:max1}) in terms of 
$n$ and $n_s$, obtaining
\begin{align}
f(\alpha,\xi) \leq 2(n-n_s)\left(2n_s+1+2\sqrt{n_s(1+n_s)}\right).
\end{align}
For a given average photon number $n$, the above function is maximized with regard to the parameter $n_s$ by choosing 
\begin{align}
n_s = \frac{n^2}{1+2n}, \label{eq:optsq}
\end{align}
and obtaining
\begin{align}
f(\alpha,\xi) \leq 2n(1+n).
\end{align}
This leads to
\begin{align}
W[|\psi_{\sf G}\rangle\langle\psi_{\sf G}|](0) \geq \frac 2\pi \exp\left\{-2n(1+n)\right\} .
\end{align}
}
\par\noindent
By looking at the proof, we remark that the bound obtained is tight: given a fixed energy $n$, by choosing the phases according to condition (\ref{eq:phases}) and
the squeezing energy according to (\ref{eq:optsq}), it is always possible to find a family of pure Gaussian states saturating the inequality. In particular the maximization obtained via condition (\ref{eq:phases}) simply corresponds, at fixed $n_d$ and $n_s$, to displace the state along the direction of the squeezed quadrature. The condition (\ref{eq:optsq}) shows that for small values of $n$, the minimum of the Wigner function is obtained by using the energy in displacement, while for larger values of $n$, the optimal squeezing fraction $n_s$ tends to an asymptotic value $n_s^{\sf (as)} =n/2$.\\
Let us now generalize the bound obtained to a generic convex mixture of Gaussian states. 
\begin{proposition}[Lower bound on the Wigner function at the origin for a convex mixture of Gaussian states]\label{p:boundmix}
For any single-mode quantum state $\varrho$ which belongs to the Gaussian convex hull 
$\mathcal{G}$, the value of the Wigner function at the origin is bounded by
\begin{align}
W[\varrho](0) \geq \frac2\pi \exp\{ -2 \bar{n} (1+\bar{n})\}
\label{eq:boundmix}
\end{align}
where $\bar{n}= \Tr[\varrho a^\dag a]$.
\end{proposition}
\proof{The multi-index $\blambda$, which labels every Gaussian state in the convex mixture $|\psi_{\sf G}(\blambda)\rangle = D(\alpha) S(\xi) |0\rangle$, contains the information about the squeezing $\xi$ and  displacement $\alpha$. We can then equivalently consider as variables $\blambda=\{n,n_s, \theta,\phi\}$. By exploiting the linearity property of the Wigner function we obtain
\begin{align}
W[\varrho](0) &= \int d\blambda \: p(\blambda) W[|\psi_{\sf G}(\blambda)\rangle\langle\psi_{\sf G}(\blambda)|](0) \nonumber \\
&\geq \frac 2\pi  \int d\blambda \: p(\blambda) \exp \{ -2n(1+n) \} \:, \label{eq:integral}
\end{align} 
where inequality (\ref{eq:boundpure}) has been used. By defining
\begin{align}
\widetilde{p}(n) = \int_0^{n} dn_s \int_0^{2\pi} d\phi \int_0^{2\pi}  d\theta \: p(\blambda),
\end{align}
which is a valid probability distribution with respect to the variable $n$, Eq. (\ref{eq:integral}) becomes\begin{align}
W[\varrho](0) \geq \frac 2\pi \int_0^{\infty} dn \: \widetilde{p}(n) \:  \exp \{ -2n(1+n)\}.
\end{align}
Studying the second derivative of 
\begin{align}
B_{\sf min}(n) = \frac 2 \pi \exp\{ -2n(1+n)\} \:,
\end{align}
we conclude that the function is {\it convex} in the whole {\em physical} region ({\em i.e.} $n\geq0$). 
As a consequence,
\begin{align}
\int_0^{\infty} dn \: \widetilde{p}(n) \: B_{\sf min} (n) \geq B_{\sf min} \left(\int_0^{\infty} 
dn\: \widetilde{p}(n)\: n \right) = B_{\sf min}(\bar{n})
\end{align}
where $\bar{n}=\int_0^{\infty} dn\: \widetilde{p}(n)\: n = \Tr[\varrho a^\dag a]$. From 
the last inequality we obtain straightforwardly the thesis
\begin{align}
W[\varrho](0) \geq \frac 2\pi \exp\{ -2 \bar{n} ( 1+\bar{n}) \} \:.
\end{align}
}
\par\noindent
The following proposition generalizes the bound obtained above.
\begin{proposition}\label{p:bound2mix} 
For any single-mode quantum state $\varrho \in \mathcal{G}$,  and for any given Gaussian map $\mathcal{E}_{\sf G}$ (or alternatively a convex mixture thereof), the following inequality holds
\begin{align}
W[\mathcal{E}_{\sf G}(\varrho)](0) \geq \frac2\pi  \exp\{ -2 \bar{n}_\mathcal{E} (1+\bar{n}_\mathcal{E})\}\:, \label{eq:bound2mix}
\end{align}
where $\bar{n}_\mathcal{E} = \Tr[\mathcal{E}_{\sf G}(\varrho) a^{\dag}a]$.
\end{proposition}
\proof{Given a quantum state $\varrho$ which can be expressed as a mixture of Gaussian state, and a Gaussian map $\mathcal{E}_{\sf G}$ (or a convex mixture thereof), the output state
\begin{align}
\varrho^\prime = \mathcal{E}_{\sf G} (\varrho)
\end{align}
can still be expressed as a mixture of Gaussian states. As a consequence we can apply to the state $\varrho^\prime$ the result in Proposition \ref{p:boundmix}, obtaining the thesis.}
\\
\par\noindent
Proposition \ref{p:bound2mix} leads  to two corollaries that will be used in the rest of the paper.
\begin{corollary}
For any single-mode quantum state $\varrho \in \mathcal{G}$,
the following inequality holds
\begin{align}
W[\varrho](\beta) \geq \frac2\pi \exp\{ -2 \bar{n}_\beta (1+\bar{n}_\beta)\}\:, \:\: \forall \beta \in \mathbbm{C} \:, \label{eq:boundDISP}
\end{align}
where $\bar{n}_\beta = \Tr[\varrho D(\beta) a^\dag a D^\dagger(\beta)]$. 
\end{corollary}
\proof{
The proof is straightforward from Proposition \ref{p:bound2mix} with the Gaussian map
$\mathcal{E}_{\sf G}(\varrho) = D(-\beta)\varrho D^\dagger(-\beta)$. We also use the 
property of the Wigner function 
$$
W[\varrho](\beta) = W[D^\dag(\beta)\varrho D(\beta)](0) \:,
$$
and $D^\dagger(\beta)=D(-\beta)$.}
\begin{corollary}
For any single-mode quantum state $\varrho$ belonging to the Gaussian convex hull 
$\mathcal{G}$,
the following inequality holds
\begin{align}
W[\varrho](0) \geq \max_{\xi \in \mathbbm C} \left( \frac2\pi \exp\{ -2 \bar{n}_\xi (1+\bar{n}_\xi)\} \right)\:,  \label{eq:boundSQ}
\end{align}
where $\bar{n}_\xi = \Tr[ \varrho S^\dagger(\xi) a^\dag a S(\xi)]$.
\end{corollary}
\proof{The proof follows from Proposition \ref{p:bound2mix}, by
considering the Gaussian map $\mathcal{E}_{\sf G}(\varrho) = S(\xi) \varrho S^\dag(\xi)$. 
Moreover, since the value of the Wigner function at the origin is invariant under any
squeezing operation, {\em i.e.}
\begin{align}
W[S(\xi)\varrho S^\dag(\xi)](0) = W[\varrho](0) \:,
\end{align}
one can maximize the rhs of 
inequality (\ref{eq:bound2mix}) with regard to the squeezing parameter $\xi$.}\\
\par\noindent
The violation of any of the inequalities presented in the last two propositions and two corollaries provides a {\it sufficient} condition to conclude that a state is quantum non-Gaussian. We formalize this by re-expressing the previous results in the form of two criteria for the detection of quantum non-Gaussianity.
\begin{criterion}\label{c:uno}
Let us consider a quantum state $\varrho$ and define the quantity
\begin{align}
\Delta_1[\varrho] = W[\varrho](0) - \frac2\pi \exp\{-2 \bar{n}(\bar{n}+1)\} \:. \label{eq:Delta1}
\end{align}
Then,
\begin{align}
\Delta_1[\varrho] < 0  \:\: \Rightarrow\:\:  \varrho \notin \mathcal{G}, \nonumber 
\end{align}
that is, $\varrho$ is quantum non-Gaussian.
\end{criterion}
\begin{criterion}\label{c:due}
Let us consider a quantum state $\varrho$, a Gaussian map $\mathcal{E}_{\sf G}$ (or a convex mixture thereof), and define the quantity
\begin{align}
\Delta_2[\varrho,\mathcal{E}_{\sf G}] = W[\mathcal{E}_{\sf G}(\varrho)](0) - \frac2\pi \exp\{-2 \bar{n}_\mathcal{E}(\bar{n}_\mathcal{E}+1)\}\:. \label{eq:Delta2}
\end{align}
Then,
\begin{align}
\exists\, \mathcal{E}_{\sf G}\:\: {\rm s.t.} \:\: \Delta_2[\varrho,\mathcal{E}_{\sf G}] < 0 
\Rightarrow \: \varrho \notin \mathcal{G}. \nonumber
\end{align}
\end{criterion}
\par
Typically, Criterion \ref{c:uno} can be useful to detect quantum non-Gaussianity of phase-invariant states having the minimum of the Wigner function at the origin of phase space. On the other hand, Criterion \ref{c:due} is of broader applicability. To give two paradigmatic examples, the latter criterion can be useful if: (i) the minimum of the Wigner function is far from the origin, so that one may be able to violate inequality (\ref{eq:boundDISP}) by considering displacement operations; (ii) the state is not phase-invariant and presents some squeezing, and thus one may be able to violate inequality (\ref{eq:boundSQ}) by using single-mode squeezing operations.
\section{Violation of the criteria for non-Gaussian states evolving in a lossy Gaussian channel}\label{s:examples}
In this section we test the effectiveness of our criteria, by applying them to typical quantum states that are of relevance to the quantum optics community. We shall consider pure, non-Gaussian states evolving in a lossy channel, and test their quantum non-Gaussianity after such evolution.  Specifically, we focus on the family of quantum channels associated to the Markovian master equation
\begin{align}\label{eq:Markov}
\frac{d\varrho}{dt} = \gamma a \varrho a^\dag - \frac\gamma 2\left(a^\dag a \varrho + \varrho a^\dag a \right).
\end{align}
The resulting time evolution, characterized by the parameter $\epsilon = 1-e^{-\gamma t}$, 
models both the incoherent loss of photons in a dissipative zero temperature environment, and 
inefficient detectors  with an efficiency parameter $\eta = 1-\epsilon$. The evolved
state $\mathcal{E}_\epsilon(\varrho_0)$ can be equivalently derived by
considering the action of a beam splitter with reflectivity $\epsilon$, which
couples the system to an ancillary mode prepared in a vacuum state.
The corresponding average photon number reads
\begin{align}
\bar{n}_\epsilon = \Tr[\mathcal{E}_\epsilon
(\varrho_0) a^\dag a] = (1-\epsilon)\: \bar{n}_0 \:,
\end{align}
where $\bar{n}_0 = \Tr[\varrho_0 a^\dag a]$ is the initial average photon number.
\\
It is well known that, for $\epsilon > 0.5$ ({\em i.e.} for detector efficiencies
$\eta<0.5$), no negativity of the Wigner function can be observed. We will focus
then on the violation of our criteria for larger values of $\epsilon$, which ensures that the evolved states have a positive Wigner function. \\
Notice that the quantum map $\mathcal{E}_\epsilon$ 
is a Gaussian map. As a consequence, by combining the divisibility property of the map (inherited from the Markovian structure of Eq.~\eqref{eq:Markov}) and Criterion \ref{c:due}, if a violation is observed for a given loss parameter $\bar{\epsilon}$, then the state is quantum non-Gaussian for any lower value $\epsilon \leq \bar{\epsilon}$ \cite{divisibility}. For this reason we will focus on the maximum values of the loss parameter $\epsilon$ for which 
a violation of the bounds is observed, {\em i.e.}
\begin{align}
\emaxU[\varrho] &= \max\{ \epsilon \: : \Delta_1[\E(\varrho)] \leq 0 \} \:, \\
\emaxD[\varrho] &= \max\{ \epsilon \: :\exists\mathcal{E}_{\sf G}\text{ s.t.} \ \Delta_2[\E(\varrho),\mathcal{E}_{\sf G}] \leq 0 \} \:.
\end{align}
\\
In what follows, we start by focusing on Criterion \ref{c:uno}, and thus we will look for negative values of the non-Gaussianity indicator $\Delta_1[\varrho]$ defined in Eq. (\ref{eq:Delta1}). 
We will consider different families of states, namely Fock states,
photon-added coherent states and photon-subtracted squeezed states.
In section \ref{s:violation2}, we will study how to improve the results obtained,
by considering the second criterion and thus by studying the 
non-Gaussianity indicator $\Delta_2 [\varrho, \mathcal{E}_{\sf G}]$. 
\subsection{Violation of the first criterion} \label{s:violation1}
\subsubsection{Fock states}
Let us start by considering Fock states $|m\rangle$, that is
the eigenstates of the number operator: $a^\dag a |m\rangle = m |m\rangle$.
A fock state evolved in a lossy channel can be written as a mixture
of Fock states as 
\begin{equation}
\mathcal{E}_\epsilon (|m\rangle\langle m|) =\sum_{l=0}^{m}\alpha_{l,m}(\epsilon)\ket{l}\bra{l}\;,
\label{eq:FockEv}
\end{equation}
with
\begin{equation}
\alpha_{l,m}(\epsilon)=
\binom{m}{l}
(1-\epsilon)^{l}\epsilon^{m-l}\;.
\end{equation}
We recall here that the Wigner function at the origin is proportional
to the expectation value of the parity operator $\Pi = (-)^{a^\dag a}$,
that is
\begin{align}
W[\varrho] ( 0)  =\frac2\pi \Tr[\varrho \Pi] = \frac2\pi \left(P_{\sf even} - P_{\sf odd} \right)\:,
\label{eq:evenodd}
\end{align}
where $P_{\sf even}$ ($P_{\sf odd}$) represents the probability of detecting
an even (odd) number of photons. By using Eq. (\ref{eq:FockEv}) one obtains 
\begin{align}
W[\E(|m\rangle\langle m|](0) = \frac2\pi (2\epsilon -1)^m \:,
\end{align}
and thus the non-Gaussianity indicator reads
\begin{align}
\Delta_1[\E(|m\rangle\langle m|] =  \frac2\pi\left\{ (2\epsilon -1)^m - e^{-2(1-\epsilon) m [ (1-\epsilon) m +1]} \right\} \:.
\end{align}
\begin{figure}[t!]
\includegraphics[width=0.495\columnwidth]{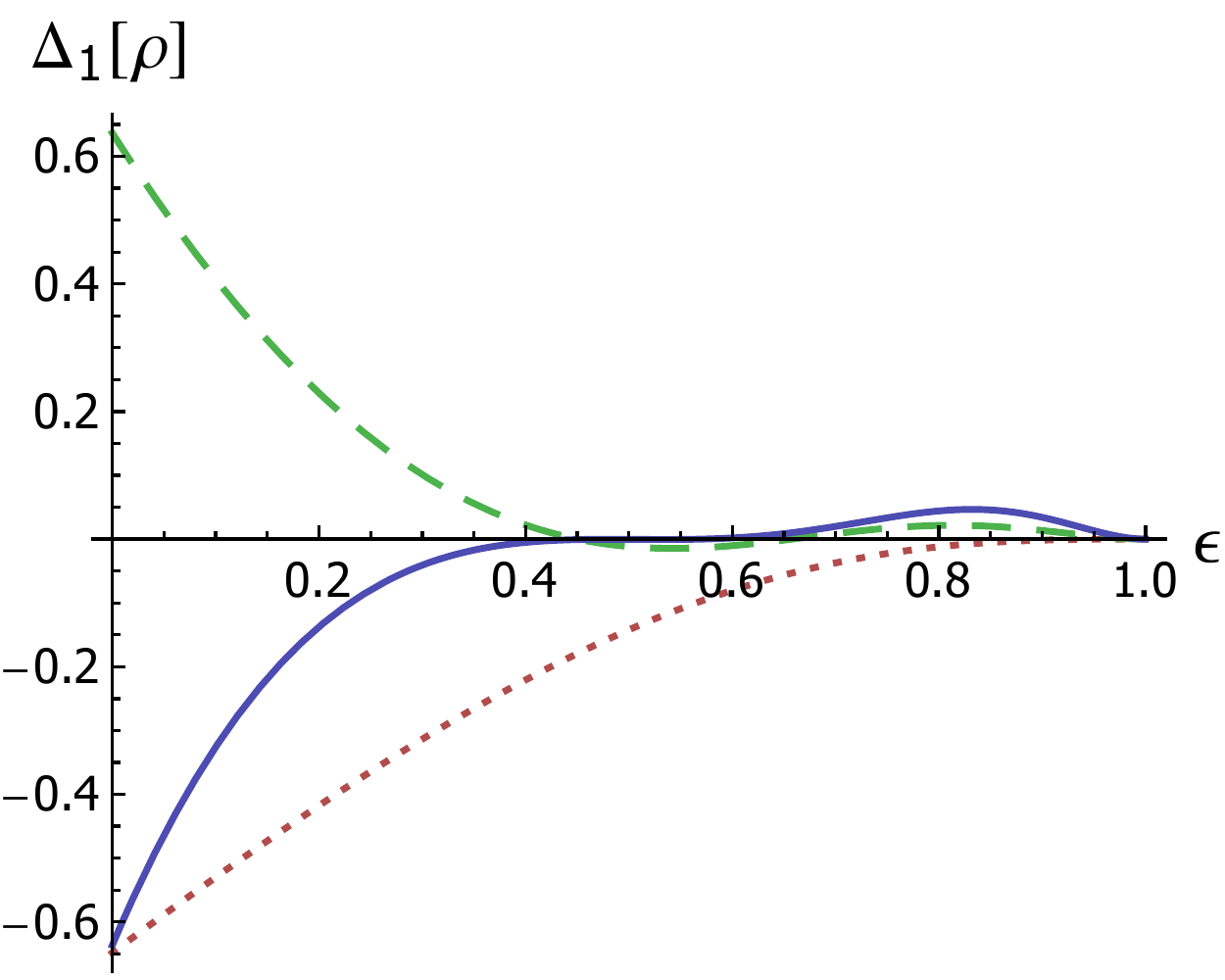}
\includegraphics[width=0.495\columnwidth]{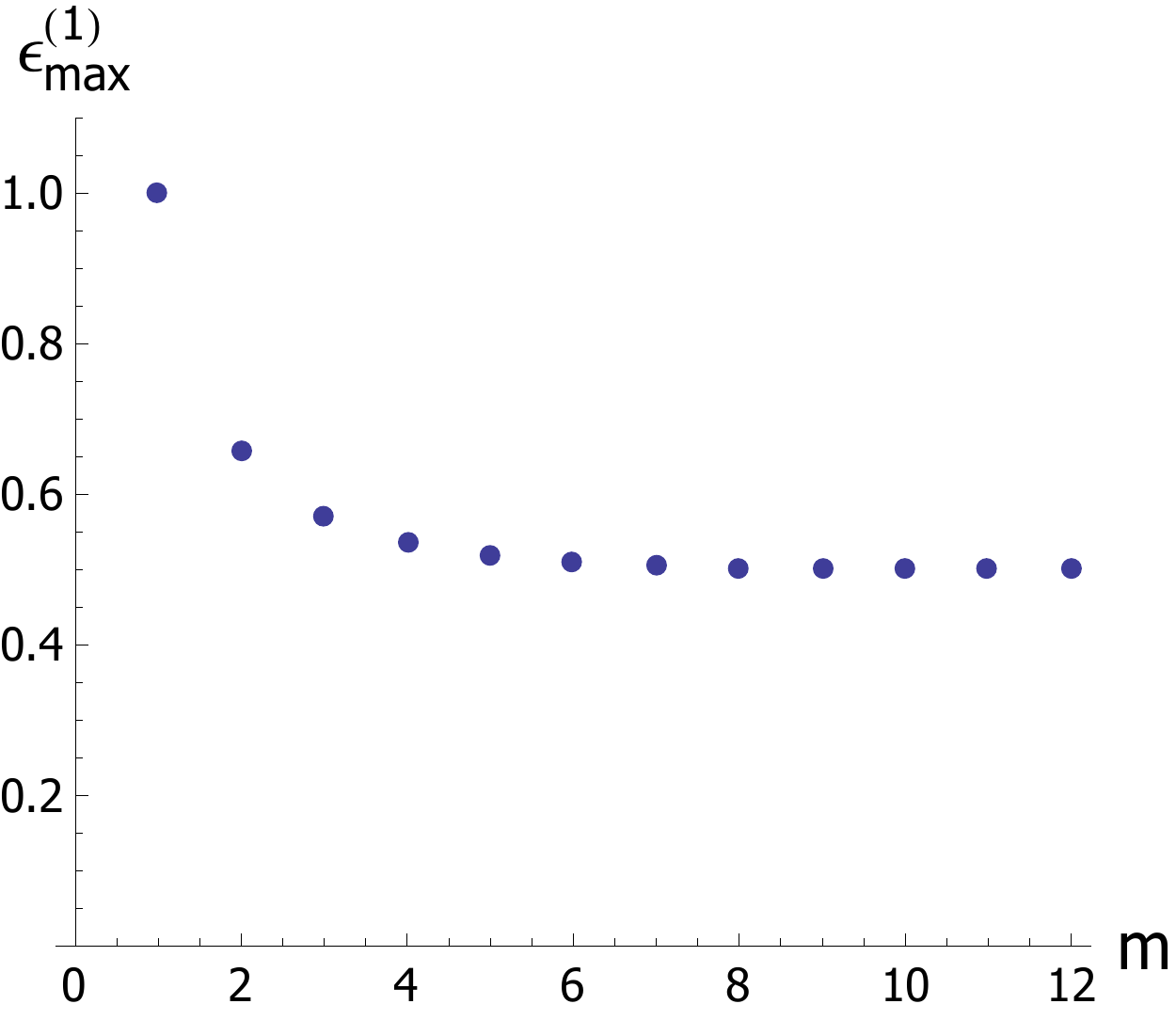}
\caption{(Left) Non-Gaussianity indicator $\Delta_1[\E(|m\rangle\langle m|)]$ for the first 
three Fock states: red-dotted line: $m=1$; green-dashed line: $m=2$; blue-solid line: $m=3$. \\
(Right) Maximum value of the noise parameter $\epsilon_{\sf max}^{(1)}$ such that the 
bound (\ref{eq:boundmix}) is violated for the state $\E(|m\rangle\langle m|)$, as a function
of the Fock number $m$.
\label{f:Fock1}}
\end{figure}
The behavior of $\Delta_1[\E(|m\rangle\langle m|]$ as a function of $\epsilon$ for the first three Fock states is plotted in Fig. \ref{f:Fock1} (left panel). 
One can observe that the criterion works really well for the 
Fock state $|1\rangle$, which is proven to be quantum non-Gaussian for all values of $\epsilon <1$. For the Fock states $|2\rangle$ and $|3\rangle$, a non monotonous behavior of $\Delta_1$ is observed as a function of the loss parameter. Still, negative values of the non-Gaussian indicator are observed in the region of interest $\epsilon>0.5$. However, the maximum value of the noise parameter $\emaxU$ decreases monotonically as a function of $m$, as shown in Fig. \ref{f:Fock1} (right panel).  By increasing the Fock number $m$, it settles to the asymptotic value $\emaxU\rightarrow 0.5$. As one would expect by looking at the bound in Eq. (\ref{eq:boundmix}), for high values of the average photon
number, the criterion becomes practically equivalent to the detection of negativity of the
Wigner function, and thus the maximum noise corresponds to $\epsilon=0.5$.
\subsubsection{Photon-added coherent states}
A \emph{photon-added coherent} (PAC) state is defined as
\begin{equation}
\ket{\psi_{\sf pac}}=\frac{1}{\sqrt{1+|\alpha|^2}}a^{\dagger}\ket{\alpha}.
\end{equation}
The operation of photon-addition has been implemented in different contexts
\cite{zavatta04,zavatta07,parigi07,zavatta09}, and in particular non-Gaussianity
and non-classicality of PAC states have been investigated in \cite{barbieri10}. \\
Being the non-Gaussianity indicator $\Delta_1[\varrho]$ phase insensitive, we can consider $\alpha \in \mathbbm{R}$
without loss of generality. The average photon number can be easily calculated
obtaining
\begin{align}
\bar{n}_0^{\sf (pac)} = \langle \psi_{\sf pac} | a^\dag a | \psi_{\sf pac} \rangle = \frac{\alpha^4 + 3 \alpha^2 + 1}{1+\alpha^2} \:,
\end{align}
while its Wigner function reads
\begin{align}
W[|\psi_{\sf pac}\rangle] (\lambda)&=\frac{2}{\pi}\frac{e^{-2(\alpha-\lambda)(\alpha-\lambda^*)}}{1+\alpha^2}\times\\
&\left(-1+\alpha^2+4|\lambda|^2-2\alpha(\lambda+\lambda^*)\right)\;.
\end{align}
The Wigner function of the state after the loss channel $\E(|\psi_{\sf pac}\rangle\langle \psi_{\sf pac}|)$ can be evaluated by means of the formula
\begin{align}
W[\E(\varrho)](\lambda)=\int \!\!d^2 \lambda' K_\epsilon(\lambda, \lambda')W[\varrho](\lambda^\prime)\;, \label{eq:WigEv}
\end{align}
where 
\begin{align}
K_ \epsilon(\lambda, \lambda')=\frac{2}{\pi\epsilon}\exp\left\{-\frac{2\left| \lambda-\lambda' \sqrt{1-\epsilon}\right|^2}{\epsilon}  \right\} \;.
\end{align}
\begin{figure}[t!]
\includegraphics[width=0.495\columnwidth]{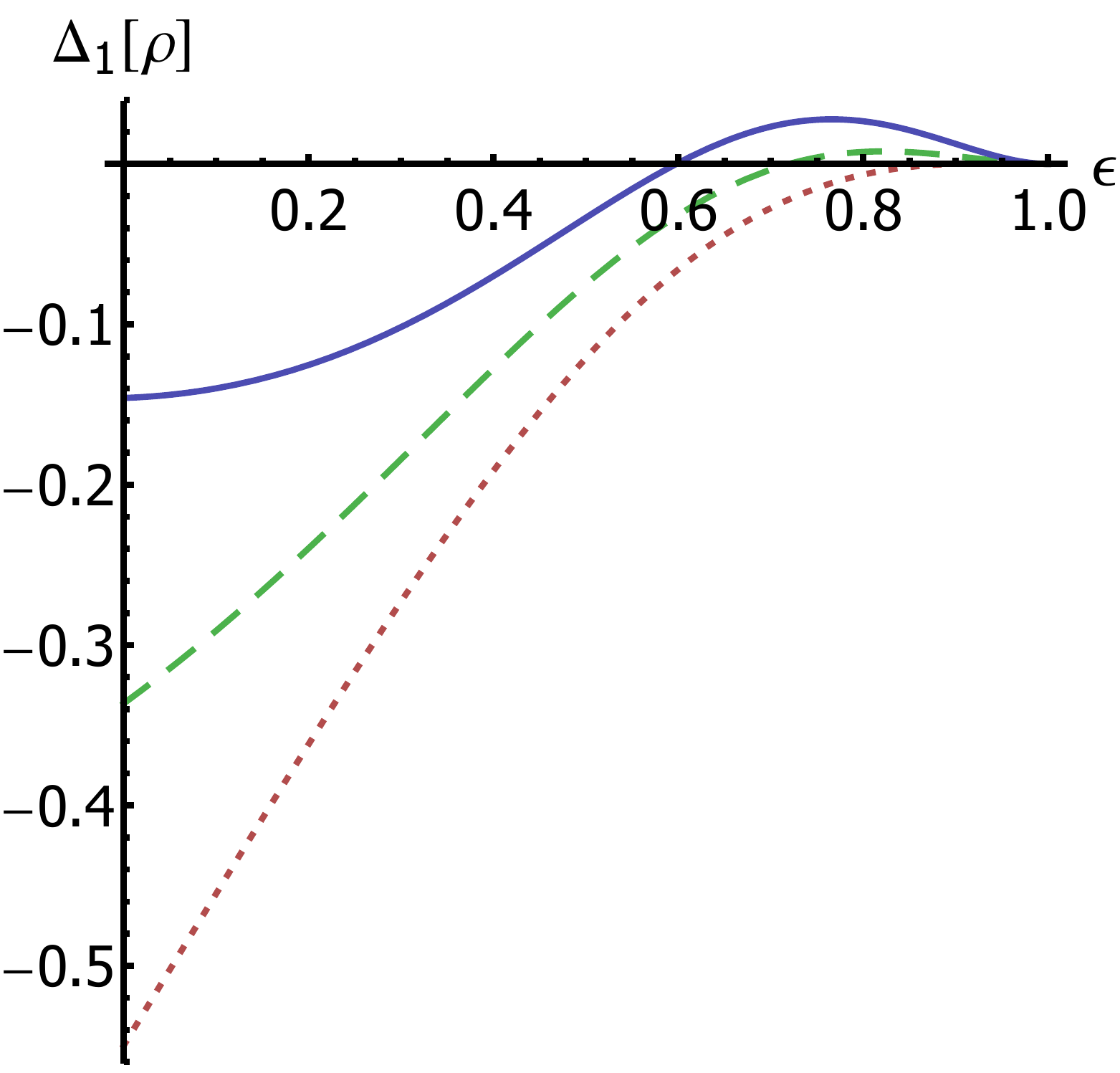}
\includegraphics[width=0.495\columnwidth]{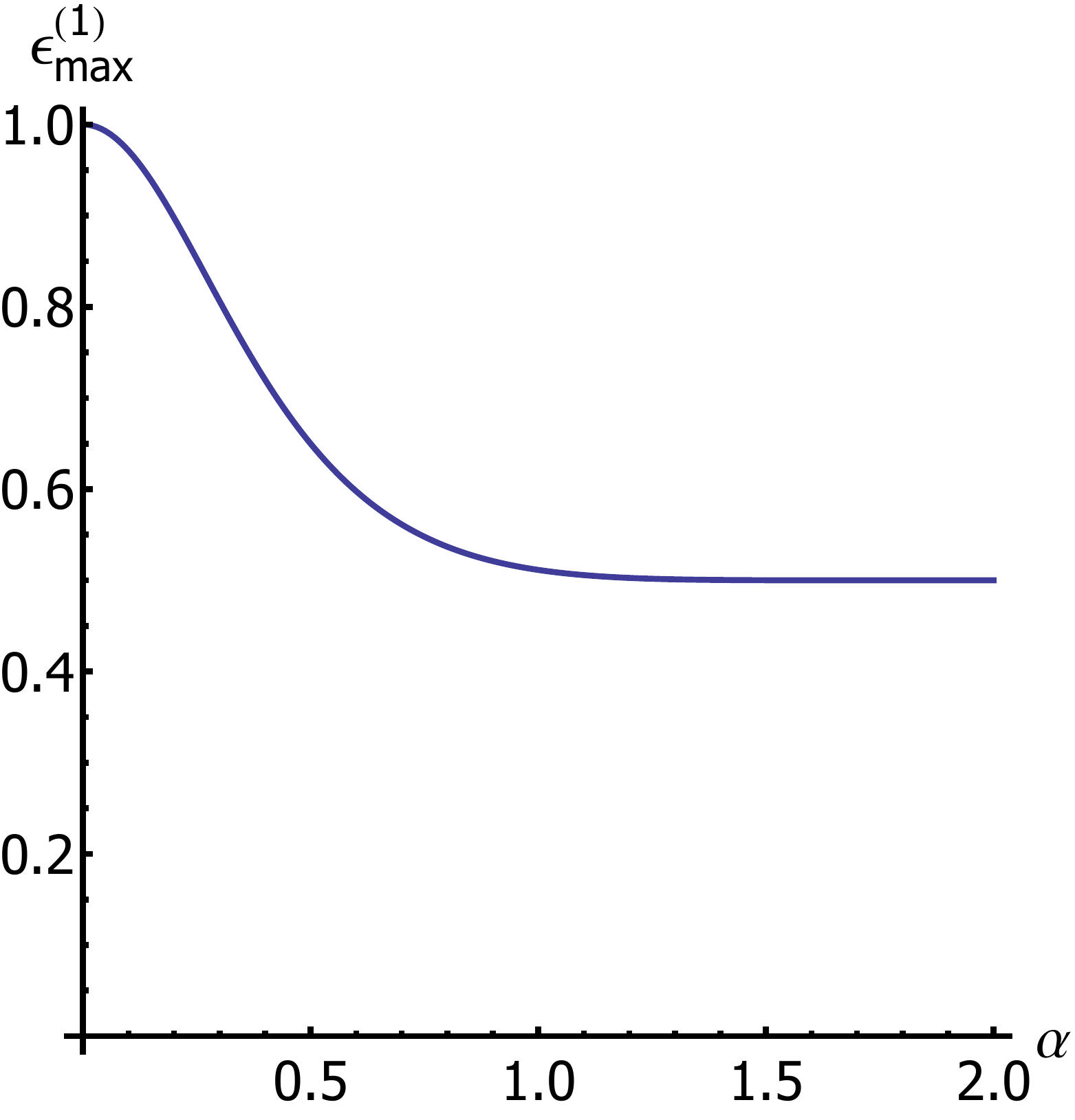}
\caption{(Left) Non-Gaussianity indicator $\Delta_1[\E(|\psi_{\sf pac}\rangle\langle \psi_{\sf pac}|)]$ for PAC states as a function of $\epsilon$ and for different values of $\alpha$: red-dotted line: $\alpha=0.2$; green-dashed line: $\alpha=0.4$; blue-solid line: $\alpha=0.6$. \\
(Right) Maximum value of the noise parameter $\epsilon_{\sf max}^{(1)}$ such that the 
bound (\ref{eq:boundmix}) is violated for the state $\E(|\psi_{\sf pac}\rangle\langle \psi_{\sf pac}|)$, as a function of the parameter $\alpha$.
\label{f:PACS1}}
\end{figure}
The non-Gaussianity indicator $\Delta_1[\E(|\psi_{\sf pac}\rangle\langle\psi_{\sf pac}|)$ can then 
be straightforwardly evaluated and is plotted in Fig. \ref{f:PACS1} (left) as a function of
$\epsilon$ for different values of $\alpha$. 
We note that negative values of the indicator can be observed in an interval for the noise parameter $\epsilon$, which decreases with the increase of $\alpha$. We can explain this feature by noting that, as $\alpha$ decreases, the PAC state approaches the Fock state $\ket{1}$: as a consequence its quantum non-Gaussianity can be more easily detected via Criterion \ref{c:uno}, in particular due to the minimum value of the Wigner function approaching the origin of the phase space. We plotted in Fig. \ref{f:PACS1} (right) the maximum value $\emaxU$ at which the violation of the bound is observed as a function of $\alpha$. Similarly to Fock states, we observe that by increasing the energy this value tends to the asymptotic value $\emaxU \rightarrow 0.5$.
\subsubsection{Photon-subtracted squeezed states}
\begin{figure}[t!]
\includegraphics[width=0.495\columnwidth]{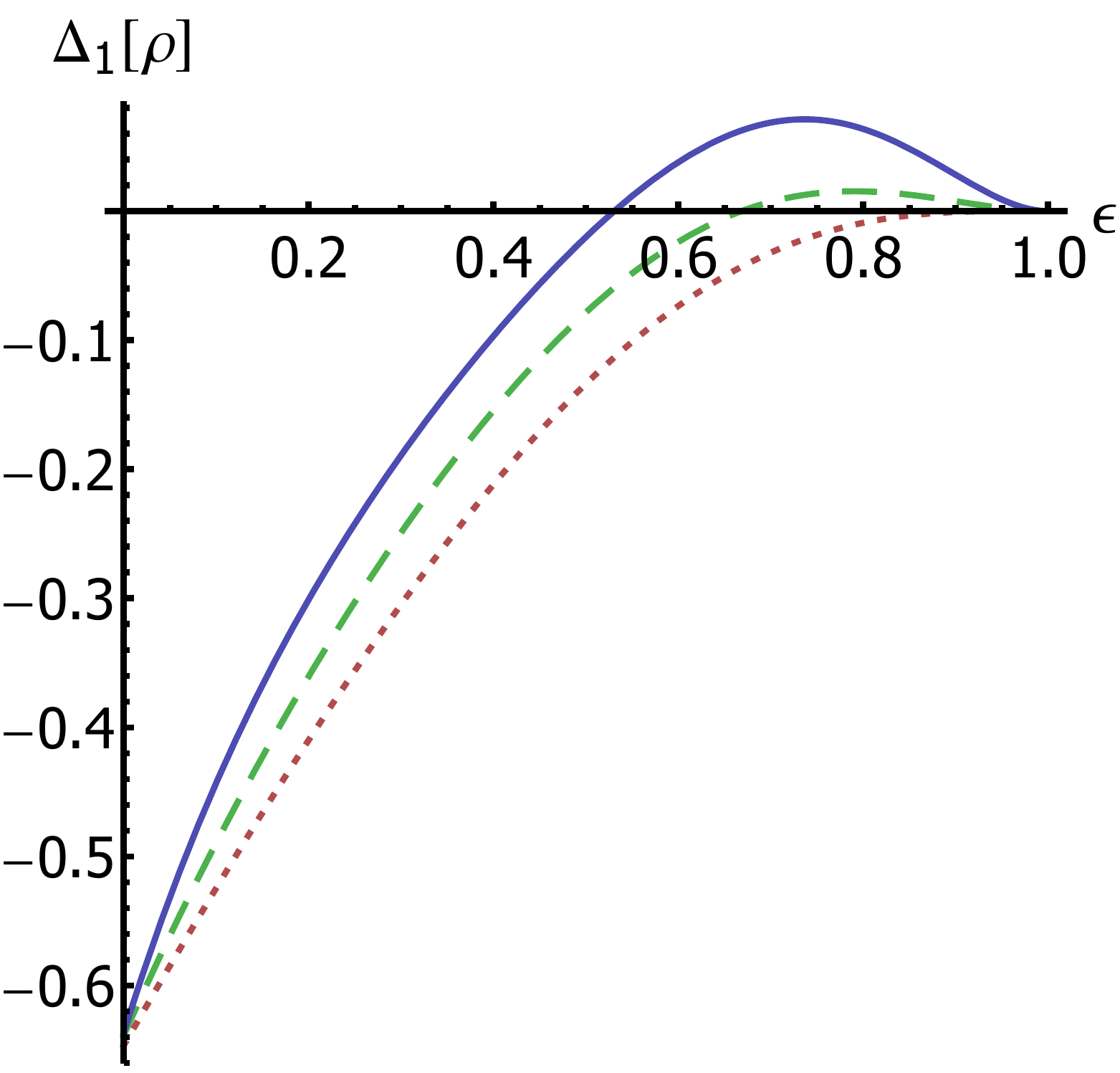}
\includegraphics[width=0.495\columnwidth]{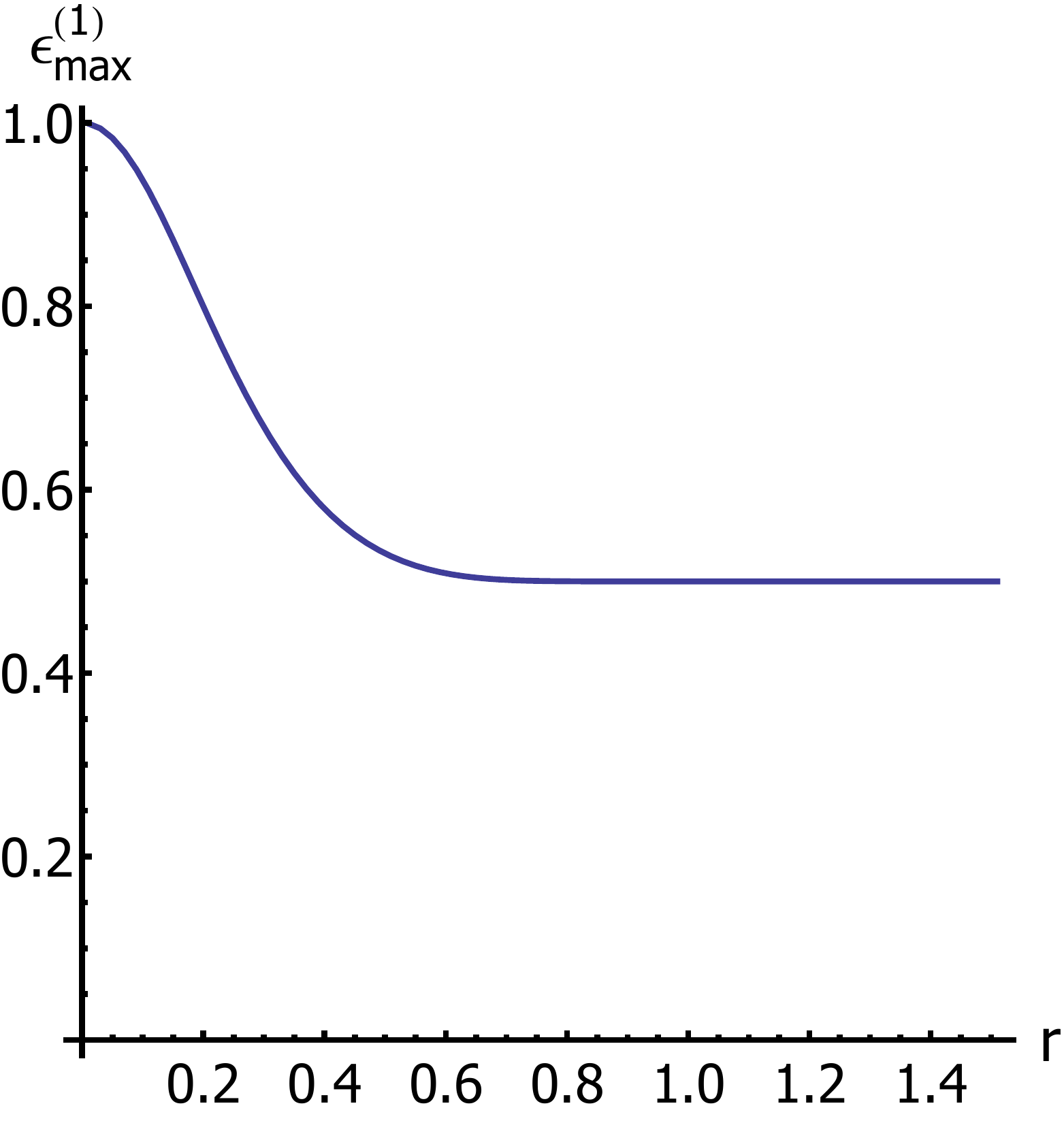}
\caption{(Left) Non-Gaussianity indicator $\Delta_1[\E(|\psi_{\sf pss}\rangle\langle \psi_{\sf pss}|)]$ for PSS states as a function of $\epsilon$ and for different values of $r$: red-dotted line: $r=0.1$; green-dashed line: $r=0.3$; blue-solid line: $r=0.5$. \\
(Right) Maximum value of the noise parameter $\epsilon_{\sf max}^{(1)}$ such that the 
bound (\ref{eq:boundmix}) is violated for the state $\E(|\psi_{\sf pss}\rangle\langle \psi_{\sf pss}|)$, as a function of the initial squeezing parameter $r$.
\label{f:PSSS1}}
\end{figure}

Let us consider now another important class of non-Gaussian states that can be engineered 
with current technology. The {\em photon-subtracted squeezed} (PSS) states are defined as
\begin{align}
\ket{\psi_{{\sf pss}}}=\frac{1}{\sinh r}aS(r)\ket{0}.
\end{align}
For low values of squeezing, these states approximate the {\em Schr\"{o}dinger kitten} states,
that is, superpositions of coherent states $|\pm \alpha\rangle$ with opposite phase and
small amplitude ($|\alpha| \lesssim 1$) \cite{dakna97}. 
The generation of this kind of states has been demonstrated experimentally \cite{ourjoumtsev06,wakui07,jonas06,gerrits10}, and it relies on performing conditional photon number measurements.\\
Without loss of generality we shall consider a real squeezing parameter $r\in\mathbbm{R}$; 
the corresponding average photon number of a PSS state reads
\begin{align}
\bar{n}_0^{\sf (pss)} = 3 \sinh^2 r +1\:,
\end{align}
while its Wigner function is
\begin{align}
W[|\psi_{\sf pss}\rangle](\lambda) &= - \frac2\pi e^{-2|\lambda|^2 \cosh{2 r} +  (\lambda^2 + \lambda^{*2}) \sinh{2 r}} \times \nonumber \\
&\left[1-4|\lambda|^2 \cosh{2 r} + 2 (\lambda^2 + \lambda^{*2}) \sinh{2 r}  \right] \:.
\end{align}
As for the PAC states, the Wigner function of the evolved state can be evaluated by means
of Eq.~(\ref{eq:WigEv}) and the non-Gaussianity indicator $\Delta_1[\E(|\psi_{\sf pss}\rangle
\langle\psi_{\sf pss}|]$ can be evaluated accordingly. Its behavior as a function of $\epsilon$ and for different values of the squeezing factor $r$ is plotted in Fig. \ref{f:PSSS1}  (left). 
In the right panel of Fig. \ref{f:PSSS1} we plot the maximum noise parameter $\emaxU$ as a function of the squeezing parameter $r$, observing the same behavior obtained for Fock and PAC states: the value of $\emaxU$ decreases monotonically with the energy of the state, approaching the asymptotic value $\emaxU
\rightarrow 0.5$.
\subsection{Violation of the second criterion} \label{s:violation2}
We will now show how the second criterion, which is based on the violation of the inequality 
(\ref{eq:bound2mix}), can be exploited in order to improve the results
shown in the previous section. Since in this case one can optimize the procedure 
over an additional Gaussian channel, in general one has $\emaxD \geq \emaxU$. The simplest Gaussian maps that one can consider are displacement and squeezing operations; correspondingly we are going to seek violation of the bounds described by Eqs. (\ref{eq:boundDISP}) and (\ref{eq:boundSQ}). As anticipated in Sec. \ref{s:criteria}, these new criteria are useful for states which are not phase invariant: the paradigmatic examples are
states {\em displaced} in the phase-space, that is, having the minimum of the Wigner function 
outside the origin, or states that exhibit squeezing in a certain quadrature. Due
to this fact, the bounds based on Eqs. (\ref{eq:boundDISP}) and (\ref{eq:boundSQ})
cannot help in optimizing the results we obtained for Fock states. We will focus then
on the other classes of states we introduced, that is PAC and PSS states. 
\begin{figure}[t!]
\includegraphics[width=0.9\columnwidth]{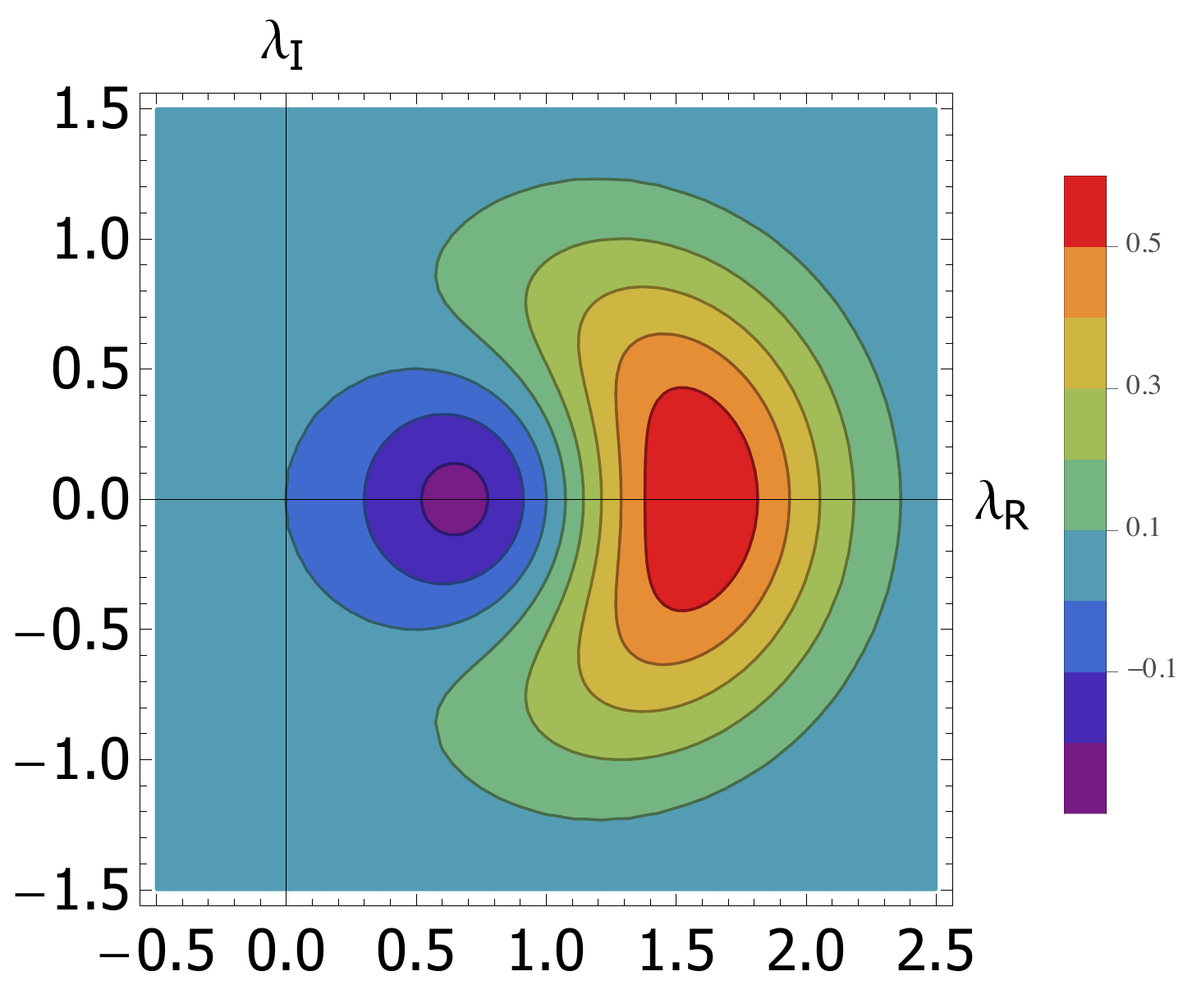}
\caption{ Contour plot of the Wigner function of the photon-added coherent state $|\psi_{\sf pac}\rangle$ for $\alpha=1$. The minimum of the Wigner function is not at the origin of the phase-space, and the state has non-zero first moments.
\label{f:WigPac0}}
\end{figure}
\subsubsection{Photon-added coherent states}
By looking at the PAC state Wigner function in Fig. \ref{f:WigPac0}, one observes that its minimum is not at the origin of the phase space. Moreover, these states have non-zero first moments, implying that one can
decrease their average photon number by applying an appropriate displacement. Both
observations suggest that it is possible to decrease the value the quantum non-Gaussianity indicator defined in Eq. (\ref{eq:Delta2}), 
\begin{align}
\Delta_{\sf pac}(\beta) = \Delta_2[\E(|\psi_{\sf pac}\rangle\langle\psi_{\sf pac}|), \mathcal{D}_\beta] \:,
\label{eq:DeltaPAC}
\end{align}
by means of a displacement operation $\mathcal{D}_\beta(\varrho) = D(\beta) \varrho D(\beta)^\dag$. To evaluate $\Delta_{\sf pac}(\beta)$ according to Eq. (\ref{eq:DeltaPAC}) one has simply to evaluate the Wigner function of the state $\varrho=\E(|\psi_{\sf pac}\rangle\langle\psi_{\sf pac}|)$ in a displaced point in the phase space, {\em i.e.} $W[\varrho](-\beta)$, and its average photon
number
\begin{align}
\bar{n}^{\sf (pac)}(\beta) &= (1-\epsilon)n_0^{\sf (pac)} + |\beta|^2 \: + \nonumber \\
& \:\: + \sqrt{1-\epsilon}( \beta^* \langle a \rangle_0 + 
\beta \langle a^\dag \rangle_0 ) \:\:,
\end{align}
where $\langle A \rangle_0 = \langle \psi_0 | A |\psi_0\rangle$, and for
$|\psi_0\rangle=| \psi_{\sf pac}\rangle$, 
\begin{align}
\langle a \rangle_0 = \langle a^\dag \rangle_0 = \frac{\alpha(2+\alpha^2)}{1+\alpha^2}\:.
\end{align}
Our goal is then to minimize $\Delta_{\sf pac}(\beta)$ 
over the possible displacement  parameters $\beta$. \\
\begin{figure}[t]
\includegraphics[width=0.495\columnwidth]{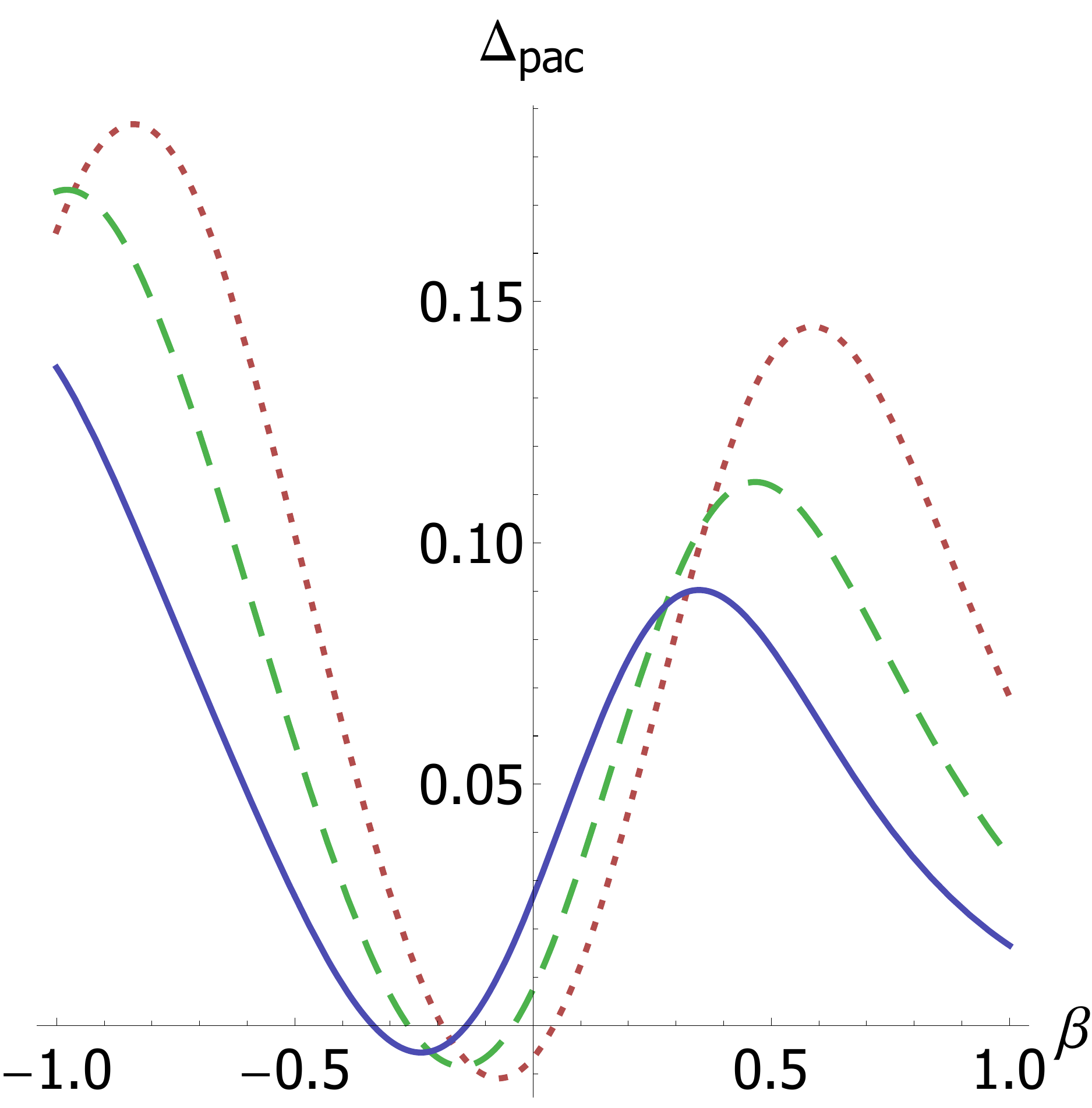}
\includegraphics[width=0.495\columnwidth]{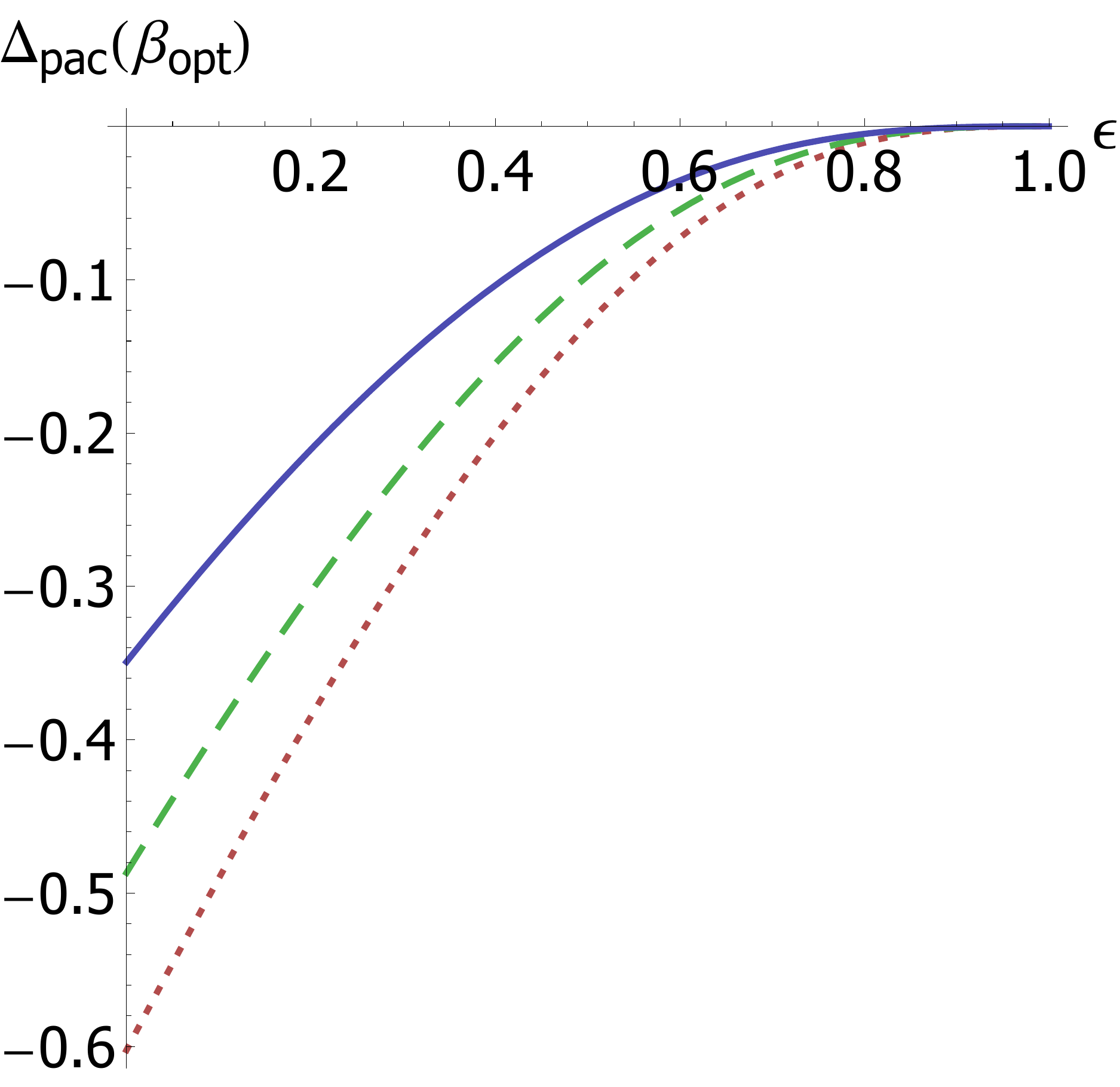}
\caption{(Left) Non-Gaussianity indicator $\Delta_{\sf pac}(\beta)$ as a function
of the additional displacement parameter $\beta$, for $\epsilon=0.8$ and for 
different values of the initial parameter $\alpha$: : 
red-dotted line: $\alpha=0.2$; green-dashed line: $\alpha=0.4$; blue-solid line: $\alpha=0.6$. \\
(Right) Optimized non-Gaussianity indicator $\Delta_{\sf pac}(\beta_{\sf opt})$ as a function
of $\epsilon$ and for different values of $\alpha$, where the displacement parameter $\beta_{\sf opt}$ has been chosen as in Eq. (\ref{eq:betaopt}):  $\alpha=0.2$; green-dashed line: $\alpha=0.4$; blue-solid line: $\alpha=0.6$. 
\label{f:PACS2}}
\end{figure}
In Fig. \ref{f:PACS2} (left) we 
plot $\Delta_{\sf pac}(\beta)$ as a function
of $\beta$ for different values of the coherent state parameter $\alpha$ and for
$\epsilon=0.8$. We observe that, while for $\beta=0$ the bound is not always 
violated, it is possible to find values such that $\Delta_{\sf pac}(\beta) <0$ and thus 
prove that the state is  quantum non-Gaussian. Unfortunately the optimal value $\beta_{\sf opt}$, which minimizes $\Delta_{\sf pac}(\beta)$, can not be obtained analytically. However we observed that for large values of $\epsilon$ 
and for $\alpha \gtrsim 1.5$ one can approximate it as
\begin{align}
\beta_{\sf opt}\simeq - \alpha\sqrt{1-\epsilon} = -\alpha e^{-\gamma t/2} . \label{eq:betaopt}
\end{align}
The behavior of  $\Delta_{\sf pac}(\beta_{\sf opt})$ as a function of $\epsilon$ shown in Fig. \ref{f:PACS2} (right), for different values of $\alpha$ and fixing $\beta_{\sf opt}$ as in Eq.~\eqref{eq:betaopt}. If we compare this with Fig. \ref{f:PACS1}, not only we observe an improvement in our capacity to witness quantum non-Gaussianity for these states, but we also see that $\Delta_{\sf pac}(\beta_{\sf opt})$ remains negative for all values of $\epsilon$. Indeed, numerical investigations seem to suggest that $\emaxD\simeq 1$ for all the possible values of $\alpha$: we indeed conjecture that any initial PAC state remains quantum non-Gaussian during the lossy evolution induced by Eq.~\eqref{eq:Markov}, and that this feature can be captured by our second criterion. However, as one can observe from Fig. \ref{f:PACS2} (right), the non-Gaussianity indicator approaches zero quite fast with both $\alpha$ and $\epsilon$, and thus it may be more challenging to detect its negativity in an actual experiment for states with a high average photon number and for large losses.
\subsubsection{Photon-subtracted squeezed states}
\begin{figure}[t!]
\includegraphics[width=0.9\columnwidth]{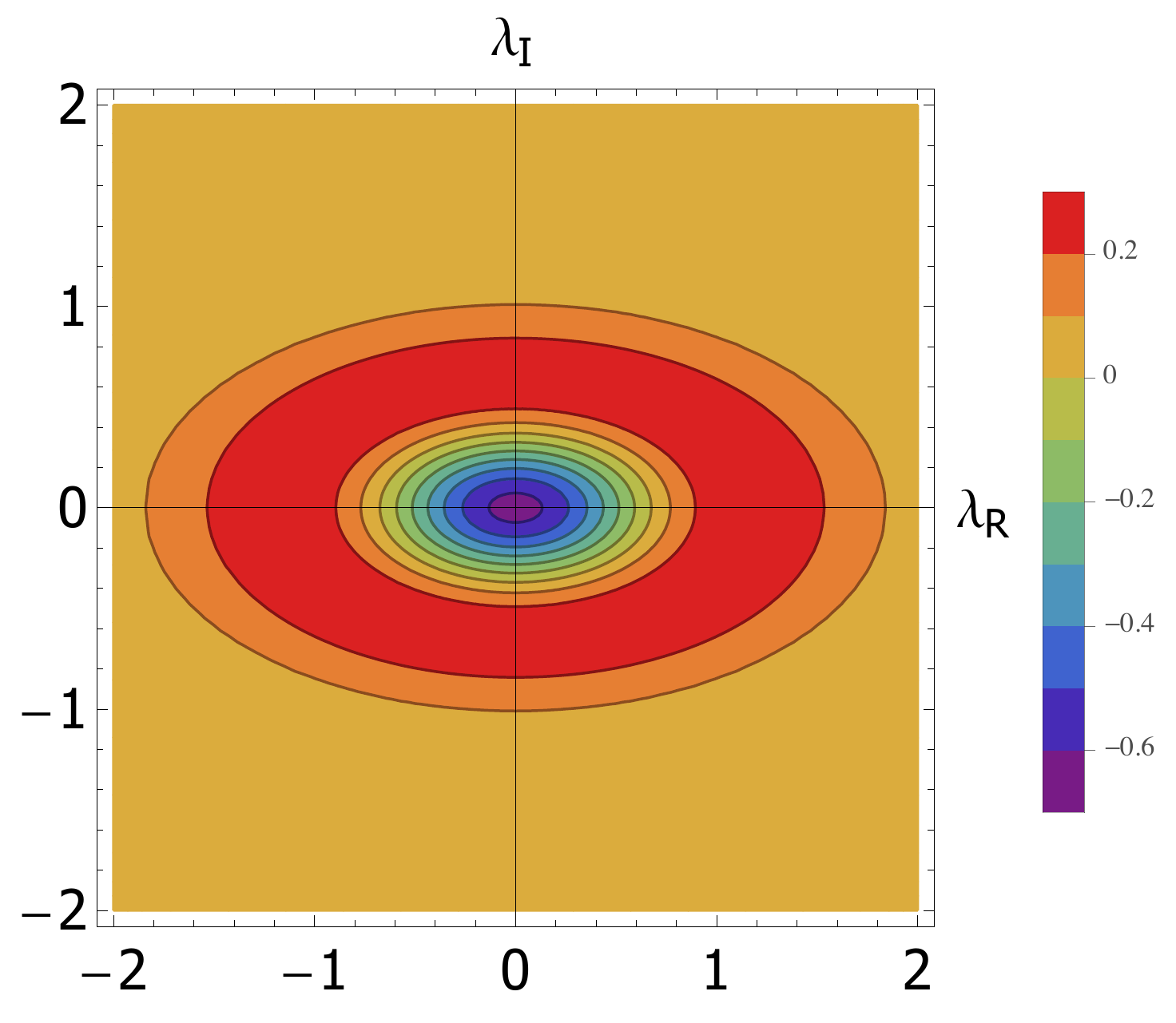}
\caption{ Contour plot of the Wigner function of the photon-subtracted squeezed state $|\psi_{\sf pss}\rangle$ for $r=0.3$. The minimum of the Wigner function is at the origin of the phase-space, and the state exhibits squeezing in one of the quadratures.
\label{f:WigPss0}}
\end{figure}
Like PAC states inherit a displacement in phase space from the initial coherent states,
PSS states inherit squeezing, as we can observe by looking at 
the Wigner function in Fig. \ref{f:WigPss0}. 
This motivates us to make use of Corollary \ref{c:due}, and thus optimize the non-Gaussianity indicator in Eq. (\ref{eq:Delta2}) as
\begin{align}
\Delta_{\sf pss}(s) = \Delta_2[\E(|\psi_{\sf pss}\rangle\langle \psi_{\sf pss}|, \mathcal{S}_s] \:, \label{eq:DeltaPSS}
\end{align}
that is by considering an additional squeezing operation $\mathcal{S}_s(\varrho) =S(s) \varrho S^\dag(s)$ on the evolved state $\varrho=E(|\psi_{\sf pss}\rangle\langle \psi_{\sf pss}|)$. As pointed out in the proof of inequality (\ref{eq:boundSQ}), 
the Wigner function at the origin is invariant under squeezing operations. Hence, the optimal
value $s_{\sf opt}$ that minimizes $\Delta_{\sf pas}(s)$ coincides with the value
which minimizes the average photon number of $\mathcal{S}_s =S(s) \varrho S^\dag(s)$,
\begin{align}
\bar{n}^{\sf (pss)}(s) &= (1-\epsilon)\left[ n_0^{\sf (pss)}  \left( \mu_s^2 + \nu_s^2 \right) \right.\nonumber \\ 
& \:\: \left. + \mu_s \nu_s\left( \langle a^2 \rangle_0 + \langle a^{\dag 2} \rangle_0 \right)
\right] + \nu_s^2\,,
\end{align}
where $\mu_t = \cosh t$, $\nu_t =\sinh t$ and for an initial PSS state (with a real squeezing parameter $r$), 
\begin{align}
\langle a^2 \rangle_0 = \langle a^{\dag 2} \rangle_0 = 3\mu_r\nu_r \:. 
\end{align}
\begin{figure}[t!]
\includegraphics[width=0.495\columnwidth]{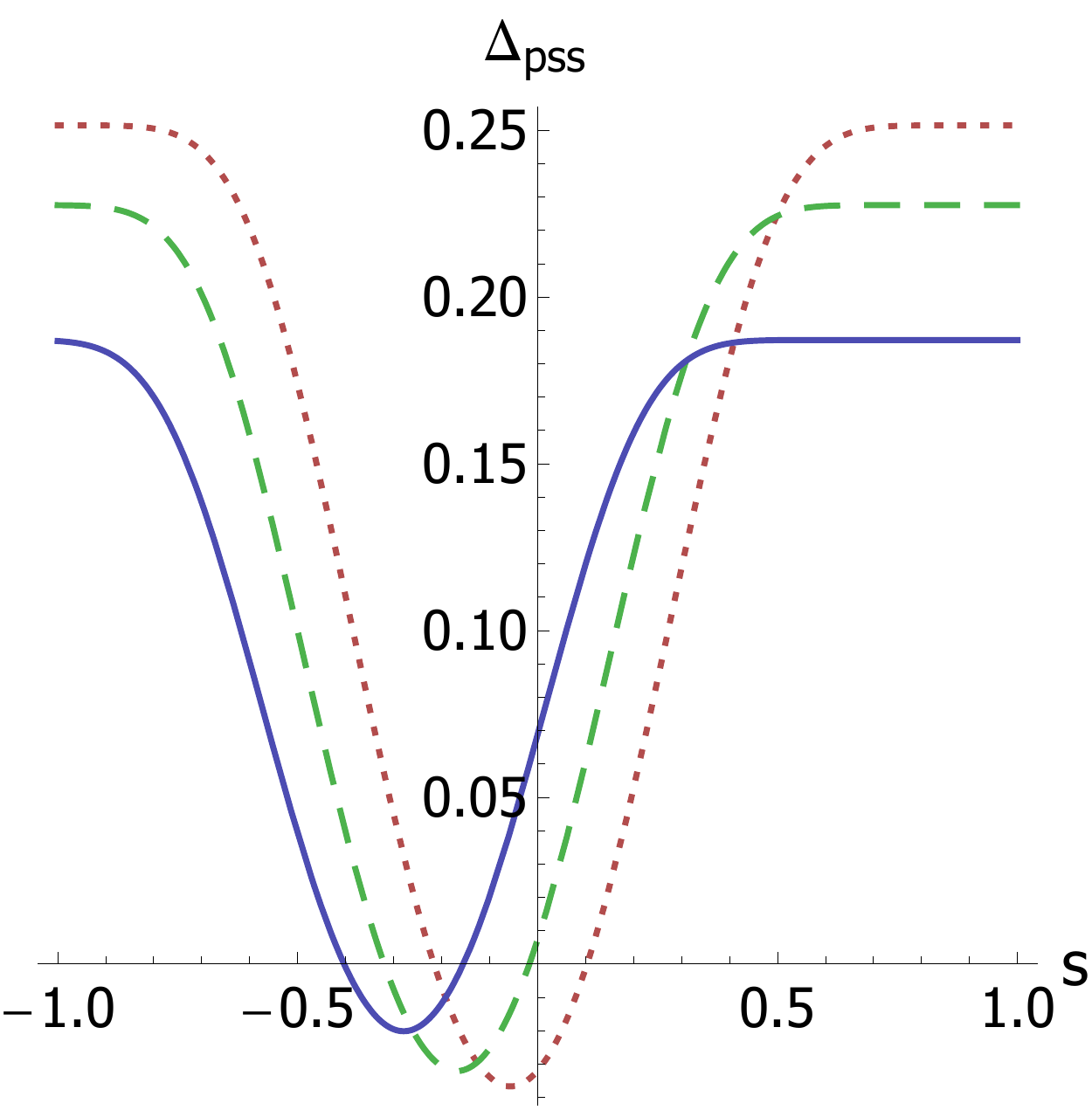}
\includegraphics[width=0.495\columnwidth]{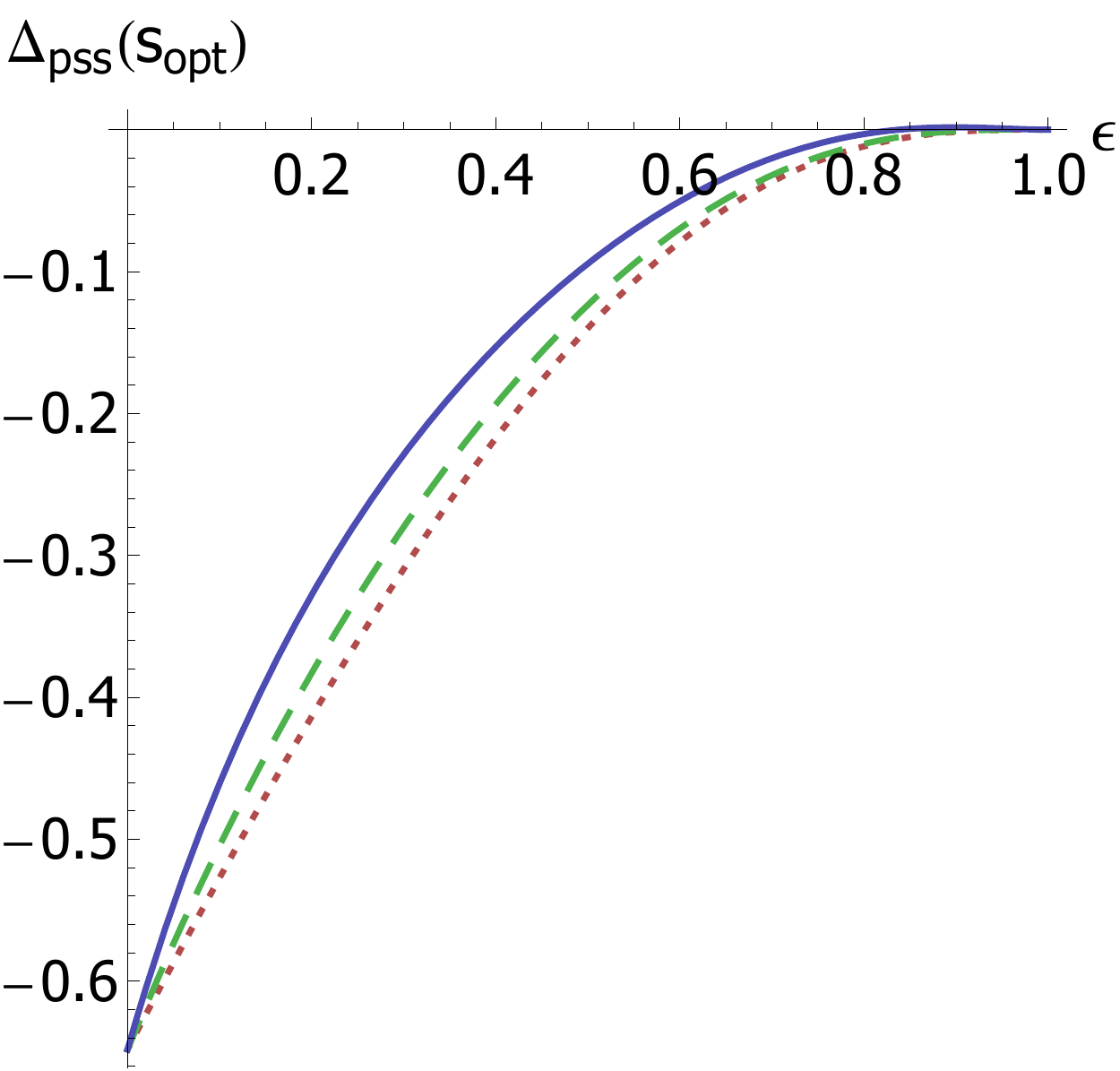}
\caption{(Left) Non-Gaussianity indicator $\Delta_{\sf pss}(s)$ as a function
of the additional squeezing parameter $s$, for $\epsilon=0.7$ and for 
different values of the initial parameter $r$: 
red-dotted line: $r=0.1$; green-dashed line: $r=0.3$; blue-solid line: $r=0.5$. \\
(Right) Optimized non-Gaussianity indicator $\Delta_{\sf pss}(s_{\sf opt})$ as a function
of $\epsilon$ and for different values of $r$, where the squeezing parameter is given by:  
red-dotted line: $r=0.1$; green-dashed line: $r=0.3$; blue-solid line: $r=0.5$. 
\label{f:PSSS2}}
\end{figure}
The behavior of $\Delta_{\sf pas}(s)$ as a function of the additional squeezing
$s$ is plotted in Fig. \ref{f:PSSS2}. As we observed in the previous case, the
optimised criterion works in cases where
the bound (\ref{eq:boundmix}) (corresponding to $s=0$) was not violated. \\
Moreover the optimal squeezing value can be evaluated analytically, 
yielding
\begin{align}
s_{\sf opt} &= - \textrm{arccosh}( \mu_{\sf opt} )\:, \\
\mu_{\sf opt}&= \frac1{\sqrt{2}} \left(1 + \frac{6 (1-\epsilon) \mu_r^2+4\epsilon-3}{\sqrt{
(4\epsilon-3)^2+12(1-\epsilon)\epsilon \mu_r^2}}\right)^{1/2} .
\end{align}
\begin{figure}[h!]
\includegraphics[width=0.8\columnwidth]{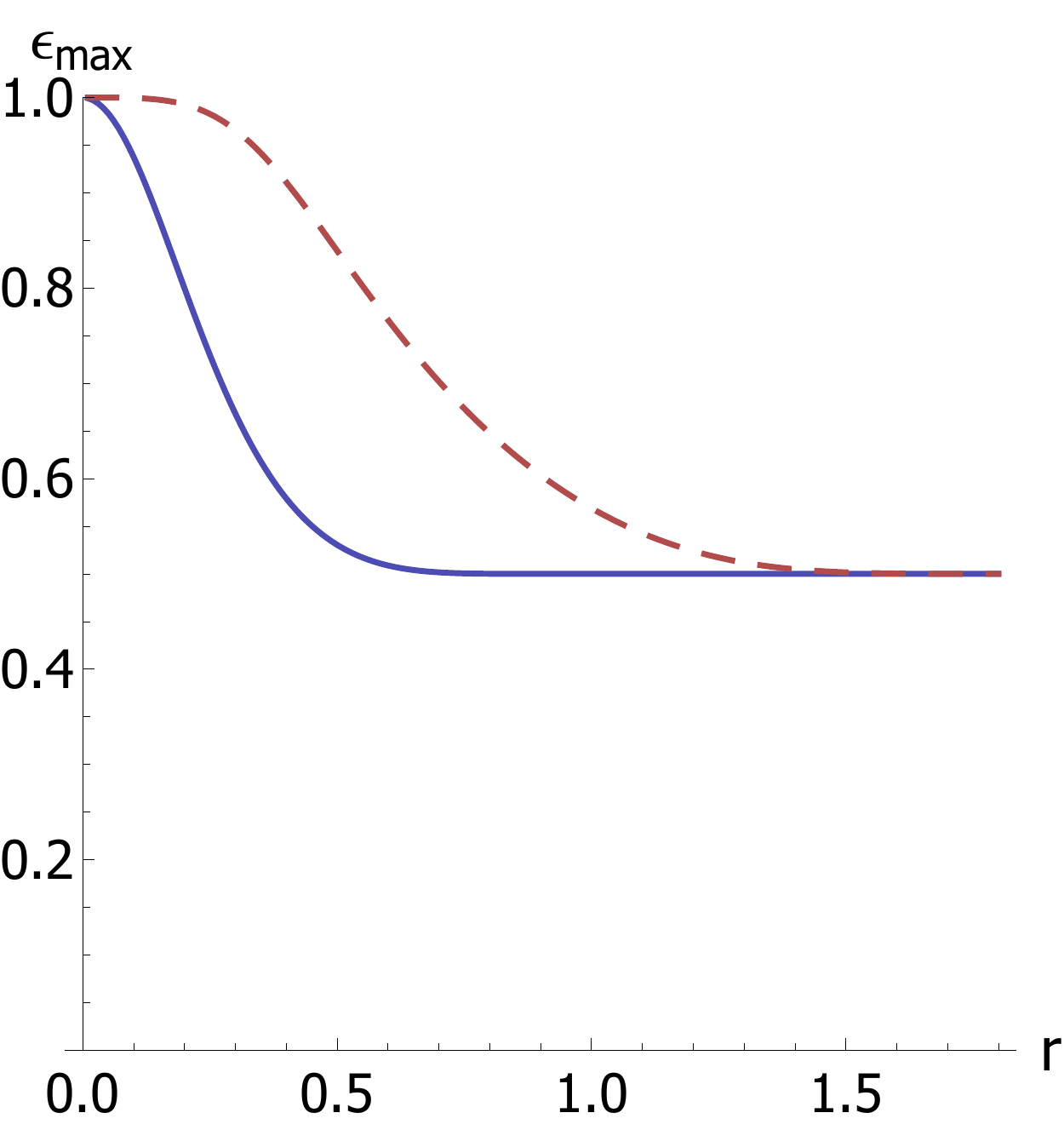}
\caption{
Maximum values of the noise parameter $\epsilon_{\sf max}^{(2)}$ and $\epsilon_{\sf max}^{(1)}$, obtained respectively by means of the optimized and not-optimized criteria, 
for the state $\E(|\psi_{\sf pss}\rangle\langle \psi_{\sf pss}|)$, as a function of the initial squeezing parameter $r$. 
Red-dotted line: $\emaxD$; blue-solid line: $\emaxU$.
\label{f:PSSS3}}
\end{figure}
The optimized quantum non-Gaussianity indicator $\Delta_{\sf pss}(s_{\sf opt})$ is 
plotted in Fig. \ref{f:PSSS2} (right), where we observe that negative values are obtained for
large values of losses. However, while for PAC states we had evidence that the 
maximum value of losses is $\emaxD\simeq 1$ for all the possible initial 
states, this is no longer true for PSS states. The behavior of $\emaxD$ as a function
of $r$ is plotted in Fig. \ref{f:PSSS3}, together with the previously obtained
$\emaxU$. We can notice the big improvement in our detection capability, obtained by exploiting Corollary~\ref{c:due}; however for large values of $r$ we still observe that  $\emaxD$ decreases towards the same limiting value $\emaxD\rightarrow 0.5$. Moreover, as it can be observed
in Fig. \ref{f:PSSS2} (right), the indicator $\Delta_{\sf pas}(s_{\sf opt})$ approaches zero
by increasing $r$ and $\epsilon$, and thus also in this case it can become challenging to witness quantum non-Gaussianity with our methods, in experiments with large values of the initial squeezing $r$ and large losses.
\section{Conclusions} \label{s:conclusions}
We have presented a set of criteria to detect quantum non-Gaussian
states, that is, states that can not be expressed as mixtures of
Gaussian states. The first criterion is based on seeking the violation of a
lower bound for the values that the Wigner function can take at the
origin, depending only on the average photon number of the state. To
verify the effectiveness of the criterion, we considered the evolution
of non-Gaussian pure states in a lossy Gaussian channel, looking for the
maximum value of the noise where such bound is violated. We observed
that the criterion works well, detecting quantum non-Gaussianity in the
non-trivial region of the noise parameters where no negativity of the
Wigner function can be observed. 
\par
We have also shown how the criterion can be generalized and improved, 
by optimising over additional Gaussian operations applied to
the states of interest. Notice that in a possible experimental
implementation one does not need to perform such additional Gaussian
operations, such as displacement or squeezing, in the actual
experiment. Indeed, it suffices to use the data obtained on the state
itself, and then apply suitable post-processing to evaluate the
optimized non-Gaussianity indicator. 
\par
Our criterion, which expresses a
sufficient condition for quantum non-Gaussianity, shares some
similarities with Hudson's theorem for pure Gaussian states, in the
sense that it establishes a relationship between the concept of
Gaussianity (combined with classical mixing), and the possible values
that a Wigner function can take. 
The successful implementation of our criteria corresponds to the
measurement of the Wigner function at the origin of the phase space 
which, in turn, corresponds to the (photon) parity of the state under 
investigation. This may be obtained with current technology
by direct parity measurement \cite{haroche07}, or by reconstruction 
of the photon distribution either by tomographic reconstruction 
or by the on/off method
\cite{Wal96,Ban96,Opa97,Mun95,Ray04,Mog98,ros04,cvp05,All09,ms07,mt03,zam04,NIST,yam99}. 
When the criterion is satisfied, one can confirm that 
the quantum state at disposal has been generated by means of a 
highly non-linear process, even in the cases where, perhaps due to 
inefficient detectors or other types of noise, negativity of the Wigner 
function can not be detected.
\section{Acknowledgments} MGG, TT and MSK thank Radim Filip for discussions. 
MGAP thanks Vittorio Giovannetti for discussions. MGG acknowledges support from UK EPSRC (EP/I026436/1). T.T. and M.S.K. acknowledge support from the NPRP 4- 426 554-1-084 from Qatar National Research Fund. SO and MGAP acknowledge support
from MIUR (FIRB ``LiCHIS'' No. RBFR10YQ3H).

\begin{thebibliography}{99}
\bibitem{Wig32} E. P. Wigner, Phys. Rev. \textbf{40}, 749 (1932)
\bibitem{Gla63} R. Glauber, Phys. Rev. {\bf 131}, 2766 (1963).
\bibitem{Sud63} E. C. G.  Sudarshan, Phys. Rev. Lett. {\bf 10}, 277 (1963). 
\bibitem{CTL91} C. T. Lee, Phys. Rev. A {\bf 44}, R2775 (1991).
\bibitem{CTL92} C. T. Lee, Phys. Rev. A {\bf 45}, 6586 (1992).
\bibitem{Arv97} Arvind, N. Mukunda, R. Simon, Phys.Rev. A {\bf 56}, 5042 (1997). 
\bibitem{Arv98} Arvind, N. Mukunda, R. Simon, J. Phys A {\bf 31}: 565 (1998).
\bibitem{Mar01} M. A. Marchiolli, V. S. Bagnato, Y. Guimaraes, B.
Baseia, Phys. Lett. A {\bf 279}, 294 (2001).
\bibitem{Par01} M. G. A. Paris, Phys. Lett. A {\bf 289}, 167 (2001).
\bibitem{Dod02} V.V. Dodonov,  J. Opt. B {\bf 4}, R1, (2002).
\bibitem{Ken04} A. Kenfack, K. Zyczkowski, J. Opt. B {\bf 6}, 396
(2004).
\bibitem{Shc05} E. V. Shchukin, W. Vogel, Phys. Rev. A {\bf 72}, 043808 (2005).
\bibitem{Kie08} T. Kiesel, W. Vogel, V. Parigi, A. Zavatta, M. Bellini, 
Phys. Rev. A {\bf 78}, 021804 R (2008).
\bibitem{Vog08} W. Vogel, Phys. Rev. Lett. {\bf 100}, 013605 (2008).
\bibitem{Sim00} R. Simon, Phys. Rev. Lett. {\bf 84}, 2726 (2000).
\bibitem{Dua00} Lu-Ming Duan, G. Giedke, J.I. Cirac, P. Zoller, Phys. Rev. Lett.
{\bf 84}, 2722 (2000).
\bibitem{Mar02} P. Marian, T. A. Marian, H. Scutaru, Phys. Rev. Lett. {\bf
88}, 153601 (2002).
\bibitem{Gio10}
P. Giorda, M. G. A. Paris, Phys. Rev. Lett. {\bf 105}, 020503 (2010);
G. Adesso, A. Datta Phys. Rev. Lett. {\bf 105}, 030501 (2010).
\bibitem{Fer12} A. Ferraro, M. G. A. Paris, Phys. Rev. Lett {\bf 108}, 260403 (2012).
\bibitem{Geh12} C. Gehrke, J. Sperling, W. Vogel, Phys. Rev. A {\bf 86}, 052118 (2012).
\bibitem{Por12} D. Buono, G. Nocerino, V. D'Auria, A. Porzio, S. Olivares, M.G.A. 
Paris, J. Opt. Soc. Am. B {\bf 27}, 110 (2010); D. Buono, G. Nocerino, A. Porzio, 
S. Solimeno. Phys. Rev. A {\bf 86}, 042308 (2012).
\bibitem{nonGHS} M. G. Genoni, M. G. A. Paris and K. Banaszek, Phys. Rev. A {\bf 76}, 042327 (2007).
\bibitem{nonGRE} M. G. Genoni, M. G. A. Paris and K. Banaszek, Phys. Rev. A {\bf 78}, 060303 (2008).
\bibitem{nonGL} M. G. Genoni and M. G. A. Paris, Phys. Rev. A {\bf 82}, 052341 (2010).
\bibitem{barbieri10} M. Barbieri, N. Spagnolo, M. G. Genoni, F. Ferreyrol, R. Blandino, M. G. A. Paris, P. Grangier and R. Tualle-Brouri, Phys. Rev. A {\bf 82}, 063833 (2010).
\bibitem{Rad11}
R. Filip, L. Mista, Jr., Phys. Rev. Lett. {\bf 106}, 200401 (2011).
\bibitem{Jez11}
M. Je\v{z}ek, I. Straka, M. Mi\v{c}uda, M. Du\v{s}ek, J.  Fiura\v{s}ek, 
R. Filip, Phys. Rev. Lett. {\bf 107}, 213602 (2011).
\bibitem{Jez12} M. Je\v{z}ek, A. Tipsmark, R. Dong, J. Fiura\v{s}ek, L.
Mi\v{s}ta, Jr., R. Filip, U. L. Andersen, Phys. Rev. A {\bf 86}, 043813 (2012).
\bibitem{predojevic12}A. Predojevic, M. Jezek, T. Huber, H. Jayakumar, T. Kauten, G. S. Solomon, R. Filip and G. Weihs, arXiv:1211.2993 [quant-ph].
\bibitem{Rad13} R. Filip, Phys. Rev. A {\bf 87}, 042308 (2011).
\bibitem{Dau10} V. D'Auria, C. de Lisio, A. Porzio, S. Solimeno, J. Anwar, M. G.
A. Paris, Phys. Rev. A {\bf 81}, 033846 (2010).
\bibitem{hudson74} R. L. Hudson, Rep. Math. Phys. {\bf 6}, 249 (1974).
\bibitem{soto83} F. Soto and P. Claverie, J. Math. Phys. {\bf 24}, 97 (1983).
\bibitem{mandilara09} A. Mandilara, E. Karpov and N. J. Cerf, Phys. Rev. A {\bf 79}, 062302 (2009).
\bibitem{titulaer65} U. M. Titulaer and R. J. Glauber, Phys. Rev. {\bf 140}, 676 (1963).
\bibitem{mari12} A. Mari and J. Eisert, Phys. Rev. Lett. {\bf 109}, 230503 (2012).
\bibitem{veitch13} V. Veitch, N. Wiebe, C. Ferrie and J. Emerson, New J. Phys. {\bf 15}, 013037 (2013). 
\bibitem{heersink06} J. Heersink, C. Marquardt, R. Dong, R. Filip, S. Lorenz, G. Leuchs and U. L. Andersen, Phys. Rev. Lett. {\bf 96}, 253601 (2006).
\bibitem{cahill69} K. E. Cahill and R. J. Glauber, Phys. Rev. {\bf 177}, 1882 (1969).
\bibitem{eisert_wolf} J. Eisert and M. M. Wolf, {\em Quantum Information with Continuous Variables of Atoms and Light}, edited by N. J. Cert, G. Leuchs and E. S. Polzik (Imperial College Press, London, 2007), pp. 23-42.
\bibitem{divisibility}
Being the map $\mathcal{E}_\epsilon$ divisible, then for all $\epsilon \leq \bar{\epsilon}$, 
a parameter $\epsilon^\prime$ exists, such that $\mathcal{E}_{\bar{\epsilon}}(\varrho) = 
\mathcal{E}_{\epsilon^\prime}(\mathcal{E}_\epsilon (\varrho) )$. As a consequence, if a criterion
is violated for the quantum state $\mathcal{E}_{\bar{\epsilon}}(\varrho)$, the quantum state
$\mathcal{E}_{\epsilon}(\varrho)$ is quantum non-Gaussian for all $\epsilon\leq \bar{\epsilon}$.
\bibitem{zavatta04} A. Zavatta, S. Viciani and M. Bellini, Phys. Rev. A {\bf 70}, 053821 (2004).
\bibitem{zavatta07} A. Zavatta, V. Parigi and M. Bellini, Phys. Rev. A {\bf 75}, 052106 (2007).
\bibitem{parigi07} V. Parigi, A. Zavatta, M. S. Kim and M. Bellini, Science {\bf 317}, 1890 (2007).
\bibitem{zavatta09} A. Zavatta, V. Parigi, M. S. Kim, H. Jeong and M. Bellini, Phys. Rev. Lett. {\bf 103}, 140406 (2009).
\bibitem{dakna97} M. Dakna, T. Anhut, T. Opatrny, L. Knoll and D.-G. Welsch,
Phys. Rev. A {\bf 55}, 3184 (1997); M. S. Kim, E. Park, P. L. Knight and H. Jeong, Phys. Rev. A {\bf 71}, 043805 (2005); S. Olivares and M. G. A. Paris, J. Opt B {\bf 7}, 616 (2005).
\bibitem{ourjoumtsev06} A. Ourjoumtsev, R. Tualle-Brouri, J. Laurat and P. Grangier,
Science {\bf 312}, 83 (2006).
\bibitem{wakui07} K. Wakui, H. Takahashi, A. Furusa and M. Sasaki, Opt. Express {\bf 15}, 3568 (2007).
\bibitem{jonas06} J. S. Neergaard-Nielsen, B. M. Nielsen, C. Hettich, K. Molmer and
E. S. Polzik, Phys. Rev. Lett. {\bf 97}, 083604 (2006).
\bibitem{gerrits10} T. Gerrits, S. Glancy, T. S. Clement, B. Calkins, A. E. Lita, A. J. Miller, A. L. Midgall, S. W. Nam, R. P. Mirin and E. Knill, Phys. Rev. A {\bf 82}, 031802 (2010).
\bibitem{haroche07} S. Haroche, M. Brune and J. M. Raimond, J. Mod. Opt. {\bf 54}, 2101 (2007).
\bibitem{Wal96} S. Wallentowitz, W. Vogel, Phys. Rev. A {\bf 53}, 4528 (1996).
\bibitem{Ban96} K. Banaszek, K. Wodkiewicz, Phys. Rev. Lett. {\bf 76},
4344 (1996).
\bibitem{Opa97} T. Opatrny and D. G. Welsch, Phys. Rev. A {\bf 55}, 1462
(1997); T. Opatrny, D. G. Welsch, and W. Vogel, Phys. Rev. A {\bf 56},
1788 (1997).

\bibitem{Mun95}M. Munroe, D. Boggavarapu, M.E. Anderson, and M. G. Raymer, Phys.
Rev. A {\bf 52}, R924 (1995); Y. Zhang, K. Kasai, and M. Watanabe, Opt. Lett.
{\bf 27}, 1244 (2002).
\bibitem{Ray04} M. Raymer and M. Beck, in {\em Quantum States
Estimation}, Lect. Not. Phys. {\bf 649} (Springer, Berlin-Heidelberg, 2004).
\bibitem{Mog98} D. Mogilevtsev, Opt. Commun. {\bf 156}, 307 (1998); Acta Phys. Slovaca {\bf
49}, 743 (1999).
\bibitem{ros04} A. R. Rossi, S. Olivares, M. G.A. Paris, 
Phys. Rev. A {\bf 70}, 055801 (2004); A. R. Rossi, M. G. A. 
Paris, Eur. Phys. J. D {\bf 32}, 223 (2005). 
\bibitem{cvp05}
G. Zambra,  A. Andreoni, M. Bondani, M. Gramegna, M. Genovese, G. Brida,
A. Rossi and M. G. A.  Paris, Phys. Rev. Lett. {\bf 95}, 063602 (2005); 
G. Zambra, M. G. A. Paris, Phys. Rev. A {\bf 74}, 063830 (2006).
\bibitem{All09}
A. Allevi, A. Andreoni, M. Bondani, G. Brida, M. Genovese, M. Gramegna, P. Traina,
S. Olivares, M. G. A. Paris, G. Zambra, 
Phys. Rev. A {\bf 80}, 022114 (2009). 
\bibitem{ms07} L. A. Jiang, E. A. Dauler, J. T. Chang, 
Phys. Rev. A \textbf{75} 062325 (2007).  A. Divochiy et al., Nat. Phot.
\textbf{2}, 302 (2008).
\bibitem{mt03} D. Achilles et al., 
Opt. Lett. {\bf 28}, 2387 (2003), M. J. Fitch, B. C. Jacobs, T. B.
Pittman, J. D. Franson, Phys. Rev. A {\bf 68}, 043814 (2003).
\bibitem{zam04} G. Zambra, M. Bondani, A. S. Spinelli, A. Andreoni, Rev. Sci. Instrum. 
{\bf 75}, 2762 (2004).
\bibitem{NIST} M. Ramilli, A. Allevi, A. Chmill, M. Bondani, M. Caccia,
A. Andreoni, J. Opt. Soc. Am. B, \textbf{27}, 852 (2010).
\bibitem{yam99} J. Kim, S. Takeuchi, Y. Yamamoto, H.H. Hogue, Appl. 
Phys. Lett. \textbf{74}, 902 (1999).
\end{thebibliography}
\end{document}